\documentclass{aa}  
\usepackage{amsmath}
\usepackage{txfonts}
\usepackage{lipsum}
\usepackage{subcaption}         
                                
\usepackage{lscape}             
                                
\usepackage{placeins}           
                                
\usepackage{tabularx}
\usepackage{graphicx}
\usepackage{CJK}
\usepackage{hyperref}
\usepackage{ulem}

{\newif\ifnotend
\notendtrue
\def\veclist{ABCDEFGHIJKLMNOPQRSTUVWXYZabcdefghijklmnopqrstuvwxyz.}
\def\top#1#2.{#1}
\def\tail#1#2.{#2.}
\loop\expandafter\xdef\csname v\expandafter\top\veclist\endcsname%
{{\noexpand\bf\expandafter\top\veclist}}
\edef\veclist{\expandafter\tail\veclist}
\if\veclist.\notendfalse\fi\ifnotend\repeat}

{\newif\ifnotend
\notendtrue
\def\callist{ABCDEFGHIJKLMNOPQRSTUVWXYZ.}
\def\top#1#2.{#1}
\def\tail#1#2.{#2.}
\loop\expandafter\xdef\csname c\expandafter\top\callist\endcsname%
{{\noexpand\cal\expandafter\top\callist}}
\edef\callist{\expandafter\tail\callist}
\if\callist.\notendfalse\fi\ifnotend\repeat}

\def\d{{\rm d}}

\def\apjs{ApJS}
\def\apjl{ApJL}

\def\aap{A\&A}

\def\Gyr{\,\mathrm{Gyr}}

\def\kpc{\,\mathrm{kpc}}

\def\kms{\,\mathrm{km\,s}^{-1}}

\def\msun{\,{\rm M}_\odot}

\begin{document}
\begin{CJK*}{UTF8}{gbsn}
\titlerunning{The multiple corrugations in the Galactic disk}
\authorrunning{J. Wang et al.}

\title{The multiple corrugations in the Galactic disk derived from the LAMOST and Gaia survey data}

%

\author{
  Jifei Wang (王吉绯)\inst{1}
  \and
  Zhuohan Li（李卓翰）\inst{2}\thanks{Corresponding authors: zhuohanli@sdu.edu.cn, chengdong.li@nju.edu.cn}
  \and
  Chengdong Li (李承东)\inst{3,4}\protect\footnotemark[1]
  \and
  Yuqin Chen (陈玉琴)\inst{1,5}
  \and
  Chengqun Yang (杨成群)\inst{6}
  \and
  Zixi Guo (郭子熙)\inst{1,5}
  \and
  Zhou Fan (范舟)\inst{1,5}
  \and
  Hongrui Gu (顾弘睿)\inst{1,5}
  \and
  Maoli Bu (布茂利)\inst{1}
}

\institute{
National Astronomical Observatories, Chinese Academy of Sciences, 100101, Beijing, People's Republic of China
\and
Shandong Key Laboratory of Space Environment and Exploration Technology, Institute of Space Sciences, School of Space Science and Technology, Shandong University, 264209, Weihai, People's Republic of China
\and
School of Astronomy and Space Science, Nanjing University, 210093, Nanjing, People's Republic of China
\and
Key Laboratory of Modern Astronomy and Astrophysics (Nanjing University), Ministry of Education, 210093, Nanjing, People's Republic of China
\and
School of Astronomy and Space Science, University of Chinese Academy of Sciences, 100049, Beijing, People's Republic of China
\and
School of Physics and Optoelectronic Engineering, Hainan University, 570228, Haikou, People's Republic of China
}

 
  \abstract
   {Large spectroscopic and astrometric surveys have revealed complex wave-like features in the Milky Way disk, suggesting that its kinematic and chemical structures are shaped by time-dependent perturbations. Recent studies have reported oscillatory patterns in the $R_{\mathrm g}-V_\phi-V_R$ space, hinting at a possible structural transition in the outer disk.}
   {We aim to characterise the transition between the inner and outer Galactic thin disk and to investigate whether radial corrugations can provide a plausible physical interpretation of the observed features.} 
   {We analysed two large stellar samples from LAMOST DR8 and Gaia DR3, combining spatial, kinematic, and chemical diagnostics. A simplified corrugation model—consisting of two radial waves propagating in opposite directions—was constructed and fitted to the observed $V_R$ pattern. We further validated the model using $N$-body simulations.}
   {Both LAMOST and Gaia samples reproduce the previously reported wave-like pattern in the $R_{\mathrm g}-V_\phi-V_R$ plane. We identify a clear transition between the inner and outer disks via the variations in rotational velocity and metallicities. The corrugation model naturally reproduces the periodic variation of $V_R$ with galactocentric radius, and the superposition of the inward- and outward-propagating modes gives rise to a comparable oscillatory pattern in both observations and simulations.}
   {Our modelling suggests that radial corrugations can provide a plausible interpretation of the observed kinematic signatures. The results highlight the complex, multi-perturber nature of the Galactic disk and motivate further investigation with upcoming surveys.}


\maketitle
\nolinenumbers

\section{Introduction}
As a typical barred spiral galaxy, the Galactic disk is one of the most prominent components of the Milky Way (MW; \citealt{binney1998,BlandHawthorn2016}). In the canonical picture of disk assembly, the MW is believed to follow an inside-out growth, with gas accretion and star formation proceeding more rapidly at small galactocentric radii ($R$) and on longer timescales in its outskirts \citep{chiappini1997chemical, chiappini2001abundance}. This process naturally leads to a negative radial metallicity gradient, which has been confirmed by large spectroscopic surveys. Based on APOGEE data, \citet{katz2021radial} found that the thin disk follows an inside-out formation up to about $10-12$ kpc, while \citet{frankel2019inside} quantified this process for the low-$\alpha$ stellar disk within 13 kpc. Nevertheless, beyond this radius, the metallicity gradient appears to flatten, and the origin of the outer disk remains under debate.

Using 70,000 red clump stars from the LAMOST survey, \citet{huang2015metallicity} found that in the solar vicinity ($7<R<11.5$ kpc), the radial metallicity gradient is moderately steep, whereas in the outer disk ($11.5<R<14$ kpc) it becomes much more gradual. \citet{chen2024inward} also mentioned that open clusters reveal a decreasing metallicity gradient in the inner disk ($R<12$ kpc) that flattens outwards. By further studying the $R-V_{\phi}$ diagram, \citet{chen2024inward} found that the slope of the ridges changes at $R \sim 11.5$ kpc; combined with the reversal of the trend in the $R_\mathrm{g}-V_{\phi}$ plane, this could be interpreted as a transition between the inner and outer disks. It has long been recognized that a transition region exists between the inner and outer thin disks  (e.g. \citealt{zhang2021radial, spina2022mapping, das2024outer, tian2024mapping}).

The theoretical interpretations in kinematic wave mechanics for stellar disks stem from the quasi-stationary density wave theory, which proposes that spiral arms are wave patterns that rotate rigidly \citep{LS1966}. On the other hand, the swing amplification, proposed by \citet{JT1966} (see also \citealt{Toomre1977,Toomre1981}), shows how transient disturbances in a differentially rotating disk could be massively amplified into prominent spiral features as they were sheared by background rotation. The wave propagation in the stellar disk and the possible mechanisms that persist the wave pattern were discussed in \citet{Toomre1969}. A crucial insight into wave maintenance was provided by \citet{LBK1972}, who demonstrated the vital role of angular momentum transfer between waves, showing how a negative angular momentum wave inside the co-rotation radius could couple with and amplify a positive angular momentum wave outside it. 

The progress in observational facilities has made significant improvements over the last few decades, including ground-based spectroscopic surveys \citep{York2000,Steinmetz2006,Yanny2009,zhao2012lamost,Majewski2017} and space astrometric telescopes \citep{HIPPARCOS,TheGaiamission}, which have helped elucidate the large-scale wave structures in the Galactic disk. Two apparent rings of stars were discovered
in the anti-center direction: the Monoceros ring \citep{Newberg2002,Yanny2003} and the Triangulum Andromeda stream, which lies at larger galactocentric radii \citep{Majewski2004,Martin2007}. Using SDSS data, \citet{Xu2015} detected asymmetry in the number counts of main-sequence stars, showing asymmetric wave patterns above and below the Galactic plane. Prominent wave-like features in both the $R-v_z$ and $R_g-v_z$ planes were discovered using Gaia TGAS data by \citet{Schonrich2018}. Similar wave-like structures are also seen in the $R-v_z$ plane \citep{Kawata2018,GaiaCollaboration2021}, the $R_g-v_z$ plane \citep{Friske2019,Antoja2022}, and the stellar energy distribution \citep{Khanna2019}. More recently, \citet{poggio2025great} presented strong observational evidence for a massive, non-axisymmetric vertical wave propagating radially outwards through the Galactic disk.

In addition to the waves in the Euclidean space, the Gaia spiral \citep{Antoja2018}, which is a non-axisymmetric structure in the phase space, indicates that the Galactic disk is out-of-equilibrium by a recent perturber. The structure includes one-armed \citep{Binney2018,Khanna2019,Frankel2025} and two-armed \citep{Hunt2022} phase spirals. The one-armed spiral is widely thought to result from the interactions between the MW and the Sagittarius dwarf galaxy (Sgr hereafter) \citep{Binney2018,Laporte2019,BlandHawthorn2021}, while the two-armed spirals probably result from internal perturbations such as the Galactic bar \citep{Banik2022,Li2023,Chiba2025}. The Gaia spiral results from perturbations to the Galactic disk, similar to those producing corrugations. However, the relation between the corrugation and the Gaia spiral remains unclear. 

Several key questions remain open in this research field, including whether multiple corrugations exist in the disk, whether the disk corrugations include both $R-\overline{V}_z$ and $R-\overline{V}_R$ patterns, how these two patterns are related, what the origins of the corrugations are, and how the corrugations are related to the warp and the Gaia phase spirals. In this paper, we investigate the multiplicity and the $R-\overline{V}_R \,\rm pattern$ of the corrugations. This paper is organized as follows. Section~\ref{sec:data} describes the two datasets adopted in this study and outlines the procedure used to select the disk samples. In Section~\ref{sec:transition}, we present the transition features between the inner and outer Galactic disks, based on multiple kinematic and chemical diagnostics. In Section~\ref{sec:discussion}, we explore a dynamical interpretation of the observational features in terms of multiple disk corrugations across the Galactic disk. Finally, we summarize the results of this study in Section~\ref{sec:conclusion}.

\section{Data}\label{sec:data}
We constructed two stellar samples based on LAMOST DR8 and Gaia DR3 data. The LAMOST sample follows the selection in \citet{chen2024inward}, targeting thin-disk stars with [Fe/H] $> -0.8$ dex and [Mg/Fe] $<$ $-0.1$ $\times$ [Fe/H] + 0.12 dex. Elemental abundances were taken from \citet{li2022stellar} catalogue, derived via a neural network (NN) trained on LAMOST DR8 and APOGEE DR17 \citep{luo2015first, zhao2012lamost, accetta2022seventeenth}. We retained stars with abundance uncertainties below 0.1 dex. Astrometric parameters (right ascension, declination, and proper motions) were obtained from Gaia DR3 \citep{TheGaiamission, GaiaDR3}, adopting a re-normalized unit weight error (RUWE) $<$ 1.4 to ensure reliable solutions \citep{lindegren2021gaia}. Radial velocities were taken from the LAMOST DR8 catalogue, with uncertainties $<$ 10 km s$^{-1}$. Distances were adopted from the \citet{anders2022photo} catalogue, derived using the StarHorse code \citep{StarHorse}, and we restricted the relative distance uncertainty to be below 30\%. To further guarantee the purity of the thin-disk population, we imposed an additional kinematic criterion of $V_\phi > 120$ km s$^{-1}$.

As a supplement and for cross-validation, we also constructed a Gaia sample using the selection of \citet{li2024exploring}. Specifically, we included stars in Gaia DR3 with available photo-astrometric distances and radial velocities \citep{anders2022photo, GaiaDR3}. The same restrictions on RUWE and relative distance uncertainty were applied as in the LAMOST sample, while the radial velocity uncertainty was limited to $<$ 20 km s$^{-1}$. As discussed in \citet{li2024exploring}, the distances of a small number of stars located behind the Galactic centre (GC) and close to the Galactic plane may be systematically overestimated due to extinction effects. To mitigate this bias, we excluded stars within the region defined by $x > 0$ kpc, $z > -3$ kpc, $r_{\rm gc} > 6$ kpc, and $-50^\circ < l < 50^\circ$, $-10^\circ < b < 15^\circ$, where $r_{\rm gc}$, $l$, and $b$ denote the galactocentric distance, Galactic longitude, and Galactic latitude, respectively.

\citet{li2024exploring} developed an NN classifier to distinguish between in situ and ex situ stellar populations. Its output is a continuous probability ranging from 0 to 1, with values close to 0 corresponding to in situ stars and values near 1 indicating ex situ stars. Based on comparisons between test particle simulations and observational data, \citet{Li2025} demonstrated that stars with predicted probabilities below 0.1 are predominantly thin-disk members. Following this method, we used the same classification threshold to initially select thin-disk candidates. In the second stage, we further refined the sample using the atmospheric parameters provided by Gaia DR3 \citep{GaiaDR3}. We required the uncertainties of [Fe/H] and $\log g$ to be less than 0.1 dex, and then selected stars with [Fe/H] $> -0.8$ dex and $\log g < 4.0$. Finally, we applied the same $V_\phi$ criterion as in the LAMOST sample.

When deriving the kinematic parameters for both samples, we adopted a circular speed at a solar radius of 232.8 km s$^{-1}$ and a solar galactocentric distance of 8.2 kpc \citep{mcmillan2016mass}. The distance from the Sun to the Galactic plane was taken as 20.8 pc \citep{Bennett19}. For the solar peculiar motion, we adopted the value reported by \citet{schonrich2010local}, yielding (U, V, W) $=$ (11.1, 12.24, 7.25) km s$^{-1}$. 

After applying all selection criteria, the final LAMOST sample contains 516,261 thin-disk stars, while the Gaia sample contains 7,306,868 thin-disk stars. In Fig.~\ref{fig:spatial_spiral_arms}, we show the spatial distribution of our sample. As we applied different selection criteria to the two samples, and since they are inherently affected by distinct selection effects and observational strategies, the two distributions exhibit noticeable differences. The Gaia sample has a much broader spatial coverage, whereas the LAMOST sample mainly extends towards the outer disk. Nevertheless, we expect the phenomenon identified in this study to appear consistently in both datasets, thereby reinforcing its physical reality.

\section{Transition features between the inner and outer disks}\label{sec:transition}
 
\begin{figure*}
 \includegraphics[width=\textwidth]{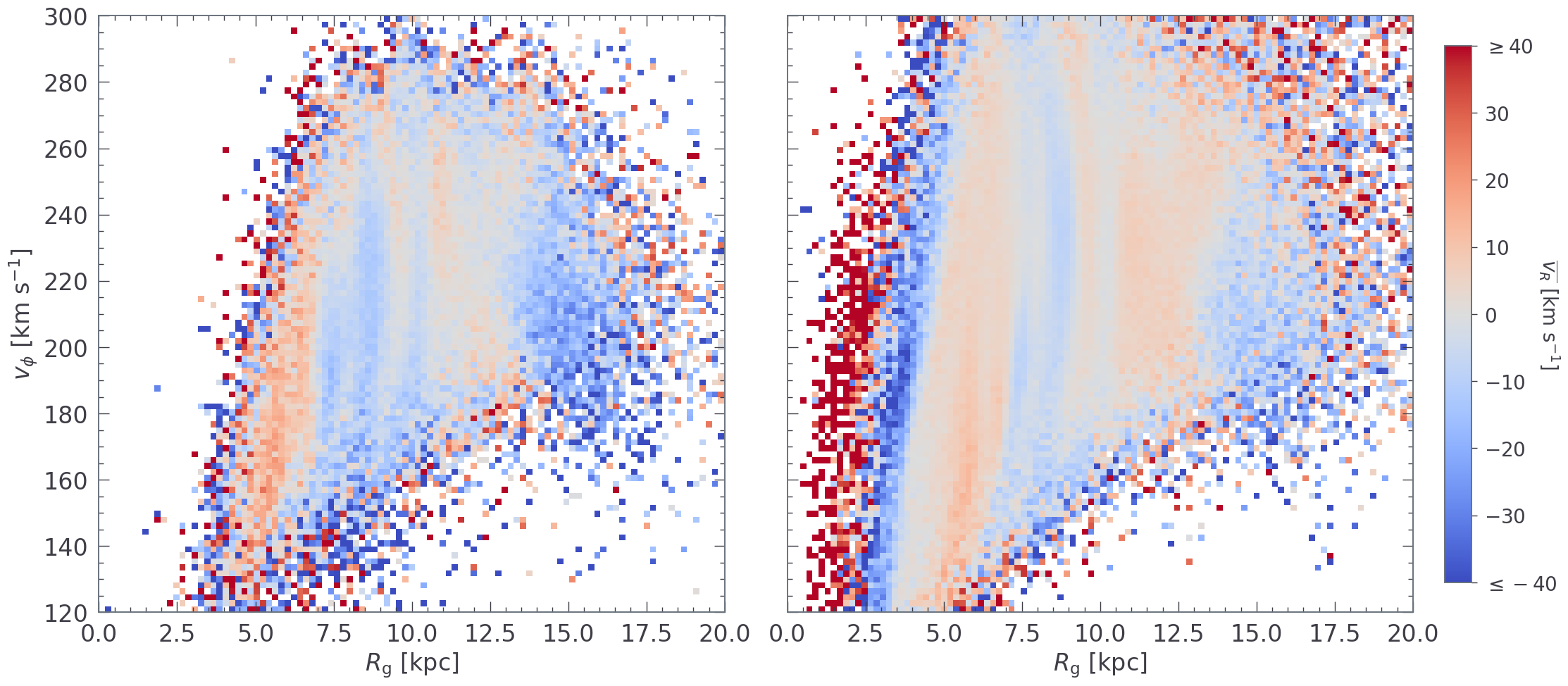}
 \centering
 \caption{Velocity waves in the $R_\mathrm{g}-V_\mathrm{\phi}-V_\mathrm{R}$ plane, following \citet{chen2024inward}. The LAMOST sample (left panel) and the Gaia sample (right panel) exhibit good consistency, both showing a clear transition at $R_\mathrm{g} \sim 13.5$ kpc.}
 \label{fig:vPhi_wave}
\end{figure*}

One of the main phenomena discovered by \citet{chen2024inward} is the presence of velocity waves in the $R_\mathrm{g}-V_\mathrm{\phi}$ plane colour-coded by $V_\mathrm{R}$, where $V_\mathrm{R}$ varies alternately with $R_\mathrm{g}$. As shown in Fig.~\ref{fig:vPhi_wave}, we reproduced this wave-like pattern using our LAMOST sample and identified a comparable structure in the Gaia data. For stars belonging to the inner part of the Galactic disk, the mean radial velocity changes direction rapidly with increasing $R_\mathrm{g}$. Although the detailed morphology of the two samples differs slightly, both display the same overall trend. In the outer part, however, the oscillation appears to vanish, giving way to a coherent inward motion in both datasets. This behaviour suggests a possible transition near $R_\mathrm{g} \sim 13.5$ kpc, separating the two kinematic regimes of the thin disk.

\begin{figure}
 \includegraphics[width=\columnwidth]{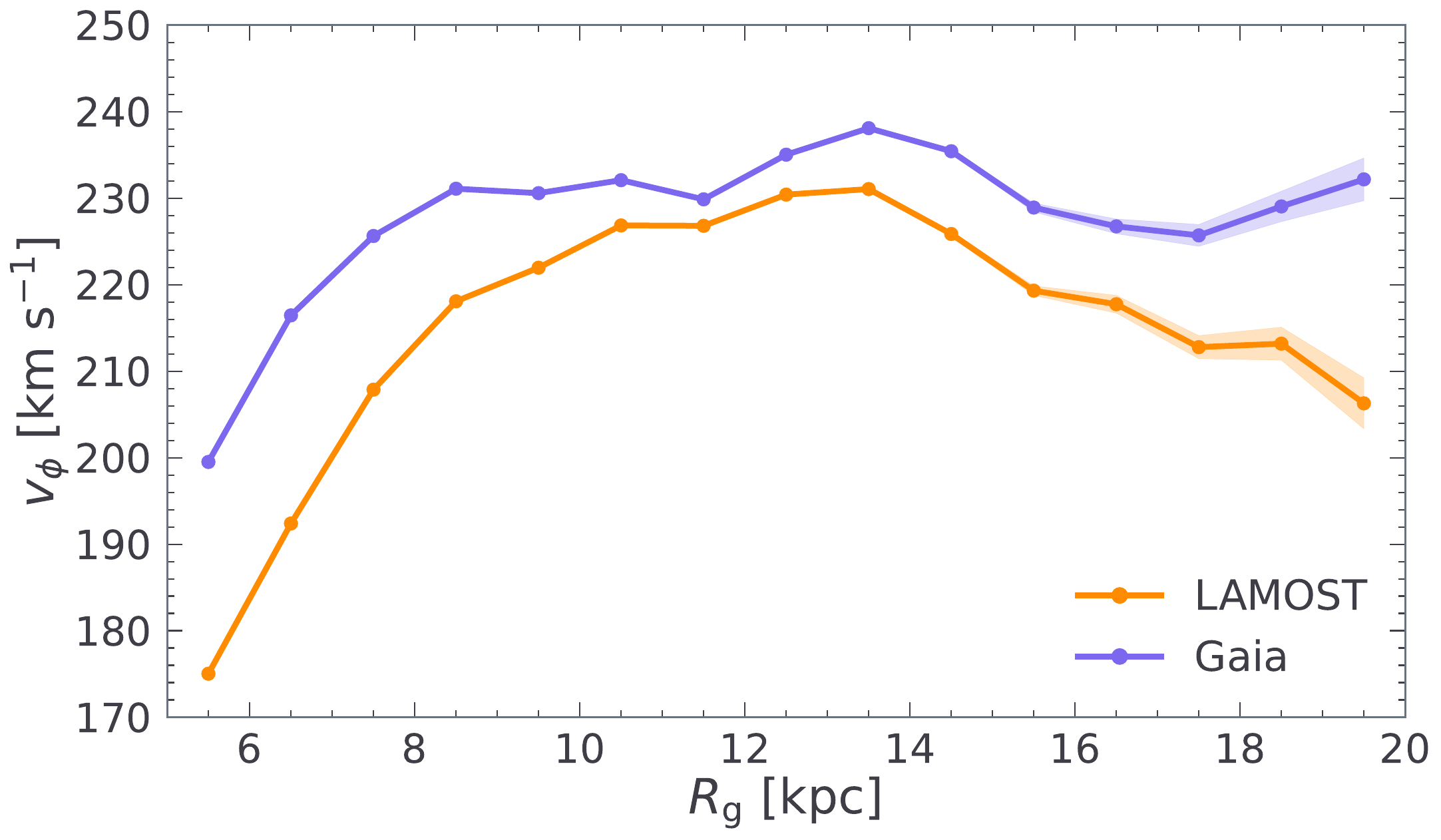}
 \centering
 \caption{Variation of rotational velocity with guiding radius. Each dot represents the arithmetic mean value within the corresponding bin, and the shaded area denotes the error of the mean. Data from LAMOST and Gaia are shown in orange and purple, respectively.}
 \label{fig:vPhi_Rg}
\end{figure}

To further investigate this transition, we examined the overall trend of rotational velocity as a function of $R_\mathrm{g}$. We divided the data into bins in $R_\mathrm{g}$ and analysed the variation of the mean rotational velocity across these intervals. As shown in Fig.~\ref{fig:vPhi_Rg}, the LAMOST sample exhibits an increase in rotational velocity followed by a decline, with a distinct transition near $R_\mathrm{g} \sim 13.5$ kpc. The outer disk region of the LAMOST sample primarily covers Galactic longitudes between $100^\circ$ and $240^\circ$. To ensure a comparable spatial coverage, we applied a similar restriction to the Gaia sample in this figure, selecting stars in the second and third Galactic quadrants ($90^\circ < l < 270^\circ$). Although the absolute values of $V_\phi$ differ between the two samples, reflecting differences in sample definition and observational characteristics, we note that the longitude restriction applied to the Gaia sample is limited to the outer disk stars, and therefore the overall spatial distributions of the two samples remain substantially different. Nevertheless, both datasets consistently show a change in the $V_\phi-R_\mathrm{g}$ trend around $R_\mathrm{g} \sim 13.5$ kpc.

This change in azimuthal velocity is not independent of the radial velocity pattern discussed above. In a non-equilibrium disk, coherent large-scale radial motions can lead to a redistribution of stars in phase space, such that the relative contributions of stars on different orbital phases vary systematically with radius. When averaged over radius, this non-equilibrium phase-space sampling can give rise to systematic variations in the mean azimuthal velocity. In this sense, the observed behaviour of $V_\phi$ provides a complementary kinematic signature that is consistent with the presence of a large-scale radial response of the disk, traced more directly by the oscillatory pattern in $V_R$. Accordingly, the inner and outer disk separated by the transition radius could be described as two kinematic regimes, reflecting different radial manifestations of the same non-equilibrium response of the disk.

The metallicity of the MW disk tends to decrease with R, providing evidence for the inside-out formation pattern of the MW disk. We find the metallicity changes similarly with $R_\mathrm{g}$. The LAMOST dataset provides a good estimation of metallicity. Gaia provides large amounts of metallicity data, which is very suitable for source selection after quality restrictions are imposed. However, it may not be very suitable for studying metallicity gradients (e.g. \citealt{andrae2023gaia}). Hence, we used the LAMOST sample alone to perform a supplementary analysis.

\begin{figure}
 \includegraphics[width=\columnwidth]{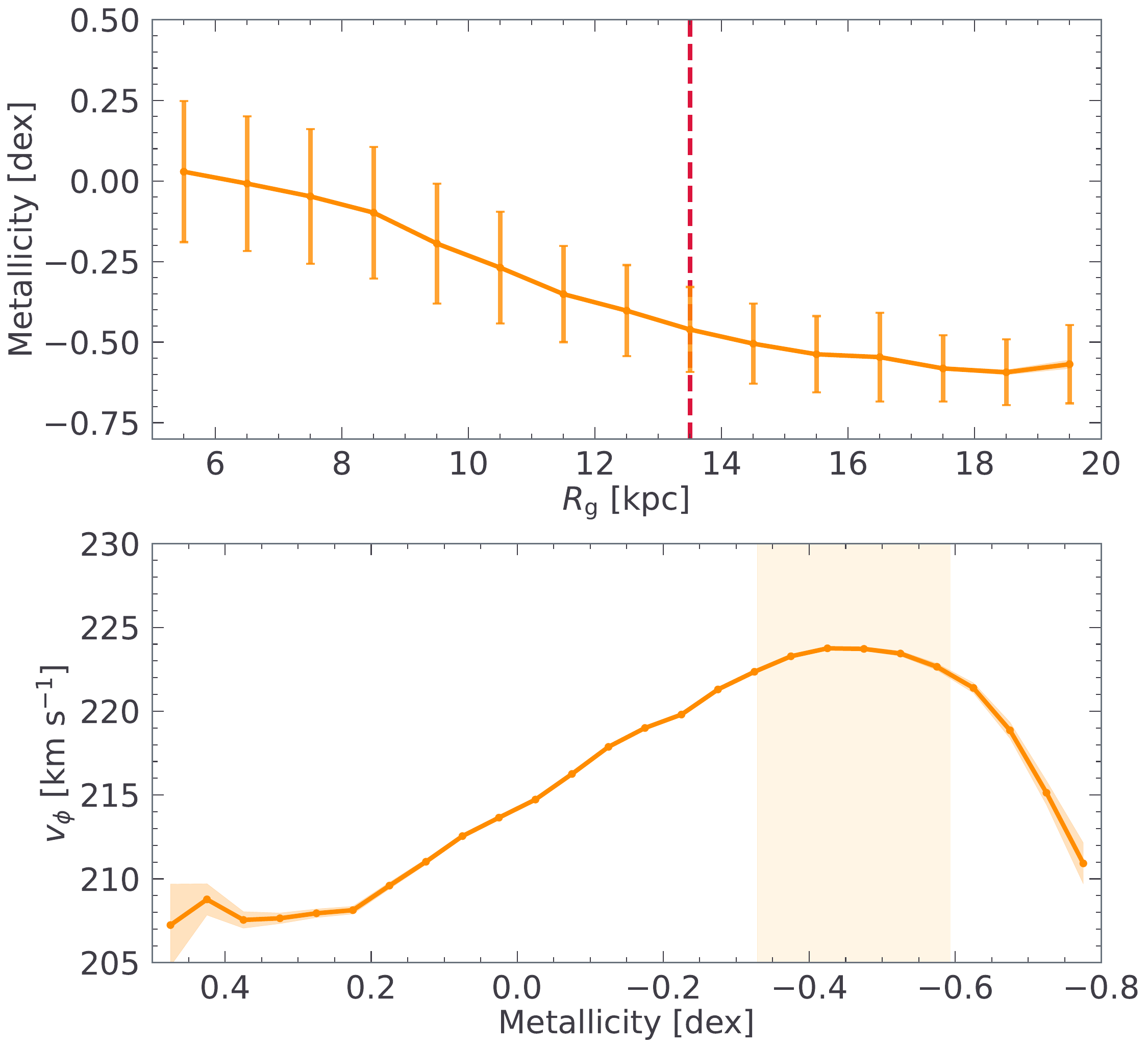}
 \centering
 \caption{Upper panel: Mean metallicity of the LAMOST sample as a function of guiding radius. Each point indicates the arithmetic mean within the corresponding bin, with the shaded region showing the uncertainty of the mean. The error bars denote the $1\sigma$ dispersion of the data in each bin. Lower panel: Rotational velocity as a function of metallicity for the LAMOST sample. The vertical shaded band marks the metallicity dispersion at $R_\mathrm{g} \sim 13.5$ kpc.}
\label{fig:vPhi_metallicity}
\end{figure}

In the upper panel of Fig.~\ref{fig:vPhi_metallicity}, we show the variation of metallicity as a function of $R_\mathrm{g}$. The trend resembles the radial metallicity gradient, displaying an overall negative slope that is steeper in the inner disk but becomes flatter in the outer region, with a transition near $R_\mathrm{g} \sim 13.5$ kpc. The lower panel illustrates an inflection in the rotational velocity, occurring at metallicities between $-0.60$ and $-0.33$ dex. This metallicity range corresponds to the interval around $R_\mathrm{g}$ $\sim$ 13.5 kpc in the upper panel. Such consistency suggests that the change in the radial metallicity trend occurs over the same radial range as the kinematic transition identified in Fig.~\ref{fig:vPhi_Rg}.

Taken together, these observational diagnostics reveal a coherent change in the properties of the thin disk around $R_\mathrm{g}$ $\sim$ 13.5 kpc. The wave-like behaviour of the mean radial velocity in Fig.~\ref{fig:vPhi_wave} provides a prominent kinematical signature of this transition, while the corresponding changes in rotational velocity and metallicity demonstrate that this feature is also imprinted in additional observable dimensions of the disk. The prominence of the radial velocity oscillations naturally points to radial dynamical processes, prompting the question of what mechanism could generate such a pattern. This consideration motivates an exploration of whether a simple radial corrugation can reproduce the observed behaviour.

\section{Multiple corrugations across the disk}\label{sec:discussion}
The observational results presented in Sect.~\ref{sec:transition} highlight the strong radial velocity oscillations that accompany the transition between the inner and outer regions of the Galactic thin disk. In this section, we investigate whether these oscillations can be interpreted within a corrugation framework. We adopt a simplified description in which the disk experiences large-scale radial displacements that vary coherently with position, producing characteristic patterns in the mean radial velocity field. The goal of this model is not to reproduce all observational trends discussed earlier, but rather to assess whether corrugations can account for the wave-like behaviour of mean radial velocity and provide a plausible dynamical ingredient relevant to the emergence of the inner-outer disk transition.

\subsection{The corrugations in the Galactic disk}\label{sec:link to corrugation}
\citet{chen2024inward} reported a group of stars moving inwards (the G1 group), which corresponds to the stellar component with $R_\mathrm{g}$ greater than 13.5 kpc in Fig.~\ref{fig:vPhi_wave}. In addition, by analysing the kinematics of two samples of young disk stars, \citet{poggio2025great} identified an outward-propagating vertical corrugation. These findings, together with our results, suggest that the transition between the inner and outer disks identified here may be related to the same underlying phenomenon revealed in previous studies, i.e. the propagation and superposition of the corrugations. 

Inspired by \citet{poggio2025great}, we constructed a corrugation model to interpret our observational results. While their formulation concentrates on vertical perturbations, our modelling focuses on the radial oscillations, since the initial signatures are traced by $V_\mathrm{R}$. Moreover, our model includes
two waves propagating in opposite directions:
\begin{equation}
\label{eq:Rhat}
\xi_{R,\mathrm{tot}}(R,t) = \xi_{R,\mathrm{out}}(R,t) + \xi_{R,\mathrm{in}}(R,t).
\end{equation}
Here, $\xi_{R,\mathrm{tot}}(R,t)$ describes the radial displacement of a star relative to its initial position. Each component in Eq.~(\ref{eq:Rhat}) follows the same functional form, which we denote generically by
\begin{align}
\xi_R(R, t) &=
A_{C}\, D_{C}(R)\,
\sin\!\left[\frac{2\pi}{\lambda_{C}}
 (R - R_{0} - v_{C} t)\right],
\label{eq:xiR_model_damped}
\end{align}
where $A_{C}$ is the amplitude of the corrugation,
and $D_{C}(R)$ is a radial damping factor, defined as
\begin{align}
D_{C}(R) \equiv \exp\!\left[-\frac{|R - R_{0}|}{L_{C}}\right],
\label{eq:D_C}
\end{align}
with $L_{C}$ being the radial damping length. The parameter $\lambda_{C}$ denotes the wavelength of the radial oscillation, $R_{0}$ is the radial offset with respect to the GC, and $t$ is the time. The propagation velocity $v_{C}$ is positive for an outward-propagating wave and negative for an inward-propagating one. In this toy model, we neglected any temporal and spatial variations of $v_{C}$ and $\lambda_{C}$ $-$ i.e. their dependence on time $t$ and galactocentric coordinates $(R,\varphi)$ $-$ in order to keep the model sufficiently simple for parameter fitting.

Correspondingly, the radial velocity was obtained by differentiating Eq.~(\ref{eq:xiR_model_damped}) with respect to time:
\begin{align}
V_R(R, t)
&= -\, A_{C}\, D_{C}(R)
   \left(\frac{2\pi v_{C}}{\lambda_{C}}\right)
   \cos\!\left[\frac{2\pi}{\lambda_{C}}
   (R - R_{0} - v_{C} t)\right].
\label{eq:vR_model_damped}
\end{align}
The radial displacement itself is difficult to trace observationally, 
whereas the corresponding oscillation in the radial velocity can be directly 
measured in both the LAMOST and Gaia samples. To avoid introducing additional 
model-dependent assumptions about the Galactic disk, we determined the parameters of the corrugation pattern empirically by fitting the observed radial velocity data.

Specifically, we binned the thin-disk sample in galactocentric radius $R$ over the range $5 \le R \le 20$ kpc into 50 equal-width bins. In each bin, the mean radial velocity was computed to form a data point in the radial velocity profile. The corrugation model in Eq.~\ref{eq:vR_model_damped} was then fitted to this binned profile using a bounded non-linear least-squares algorithm.

\begin{figure}
 \includegraphics[width=\columnwidth]{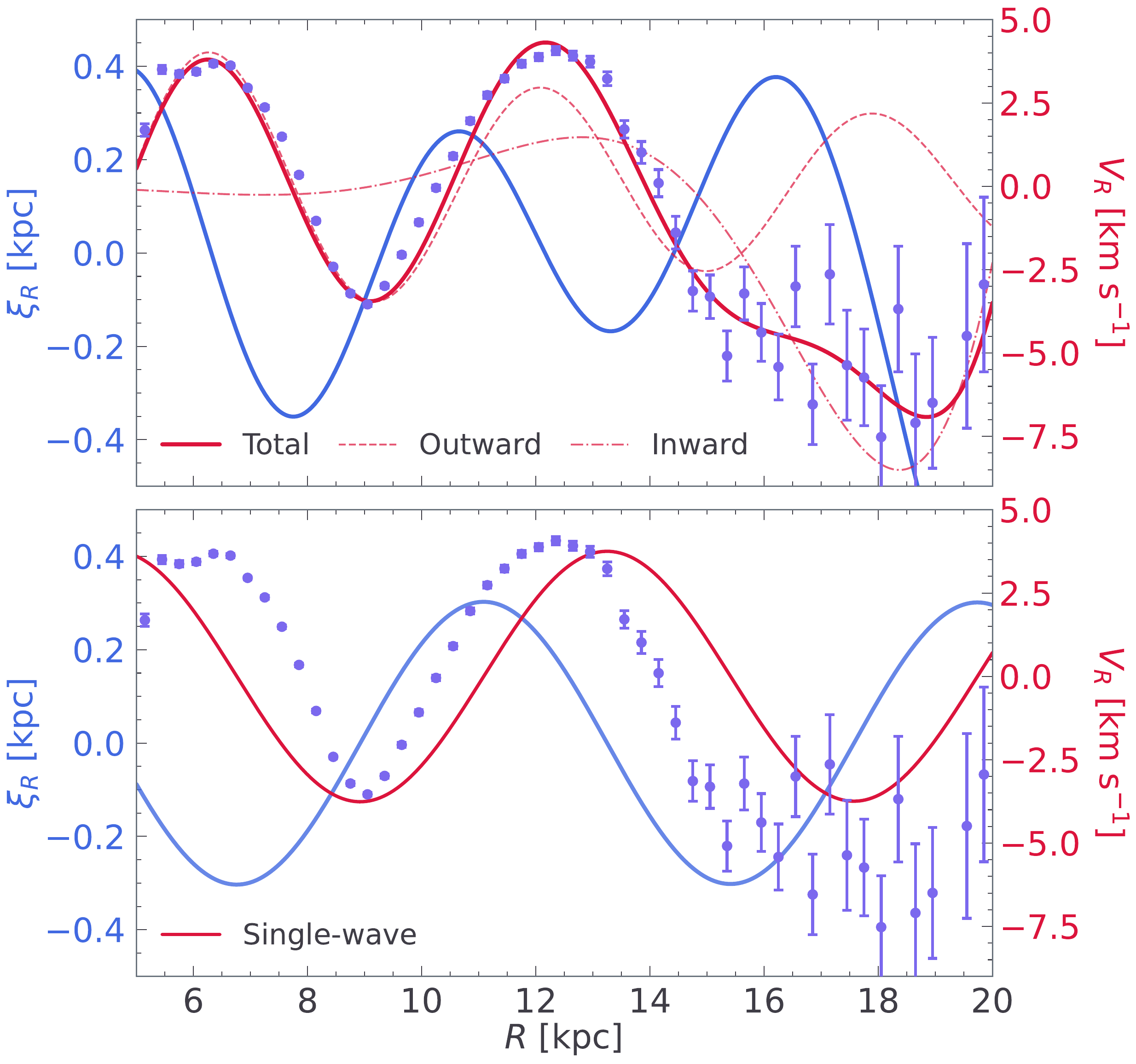}
 \centering
 \caption{Corrugation model fitted to the Gaia sample. Upper panel: Best-fitting double-wave model. The purple points show the mean radial velocity in each radial bin, with error bars indicating the uncertainty of the mean. The red curve represents the best-fitting $V_\mathrm{R}$ pattern. The inferred radial displacement $\xi_R(R)$ is shown in blue. Lower panel: Same as the upper panel, but using a single-wave model, which fails to reproduce the full structure of the observed $V_\mathrm{R}$ pattern.}
 \label{fig:radial_corrugation}
\end{figure}

The upper panel of Fig.~\ref{fig:radial_corrugation} shows the fitted radial corrugation pattern and its associated radial velocity. At the snapshot of t=0, we applied the fitting procedure to the Gaia sample and derived the corresponding radial displacement. The Gaia sequence provides a clean and well-defined trend that makes it well suited for illustrating the modelling procedure. In principle, the same method can be applied to other datasets such as LAMOST, although the Gaia data yield a more stable fit for demonstration purposes. For comparison, we illustrate the behaviour of a single wave model in the lower panel of Fig.~\ref{fig:radial_corrugation}, which demonstrates that a single wave is insufficient to reproduce the observed $V_R(R)$ pattern. The best-fitting parameters of the double wave model are listed in Table~\ref{tab:parameters}.

In Fig.~\ref{fig:vPhi_wave}, the radial velocity 
exhibits a periodic variation with guiding radius $R_{\mathrm{g}}$. A similar oscillation is visible in Fig.~\ref{fig:radial_corrugation}, but with a lower frequency and amplitude. This reduction arises because stars with different guiding radii contribute to the same radial position $R$, leading to spatial averaging that smooths the signal. Even after this smoothing, the model retains a reversal in the sign of $V_\mathrm{R}$ around 13$-$14 kpc, matching the location of the transition identified in the observations. The associated radial displacement, shown by the blue curve, reveals systematic outward and inward shifts whose peaks and troughs mark where the corrugation is most pronounced. These coherent radial motions illustrate how a large-scale corrugation can generate the wave-like behaviour of $V_\mathrm{R}$. When combined with the underlying density structure of the Galactic disk, such corrugations may contribute to shaping the large-scale features associated with the emergence of the inner-outer disk transition described in Sect.~\ref{sec:transition}.

\begin{table*}
\begin{center}
    \caption{Best-fitting parameters of the inward and outward corrugation modes.}
    \begin{tabular*}{\textwidth}{@{\extracolsep{\fill}}lccccc}
        \hline
        Mode & $A_{C}$ (kpc) & $\lambda_{C}$ (kpc) & $R_0$ (kpc) & $L_{C}$ (kpc) & $v_{C}$ (kpc Gyr$^{-1}$) \\
        \hline
        Inwards  & 2.2 & 11.1 & 23.0 & 3.2  & -33.6 \\
        Outwards & 0.4 & 5.8  & 3.4  & 19.1 & 10.2    \\
        \hline
    \end{tabular*}
\label{tab:parameters}
\end{center}
\end{table*}

Our analysis offers an alternative interpretation in line with the findings of \citet{poggio2025great}, although a full quantitative framework will require dedicated dynamical modelling. Identifying the physical drivers of the inward- and outward-propagating modes remains an open challenge. Nevertheless, the transition between the inner and outer parts of the disk revealed in this work stands out as a genuine structural signature of the MW. Understanding its origin will provide a key benchmark for future models of the Galaxy's dynamical evolution.

\subsection{Validation of the radial corrugation model with N-body simulations}
To assess whether such a corrugation can arise in a realistic dynamical environment, we further examined an N-body simulation of a MW-like disk. The details of the settings and methods of the N-body experiment are shown in 
Appendix~\ref{sec:Nbody}

\begin{figure}
 \includegraphics[width=\columnwidth]{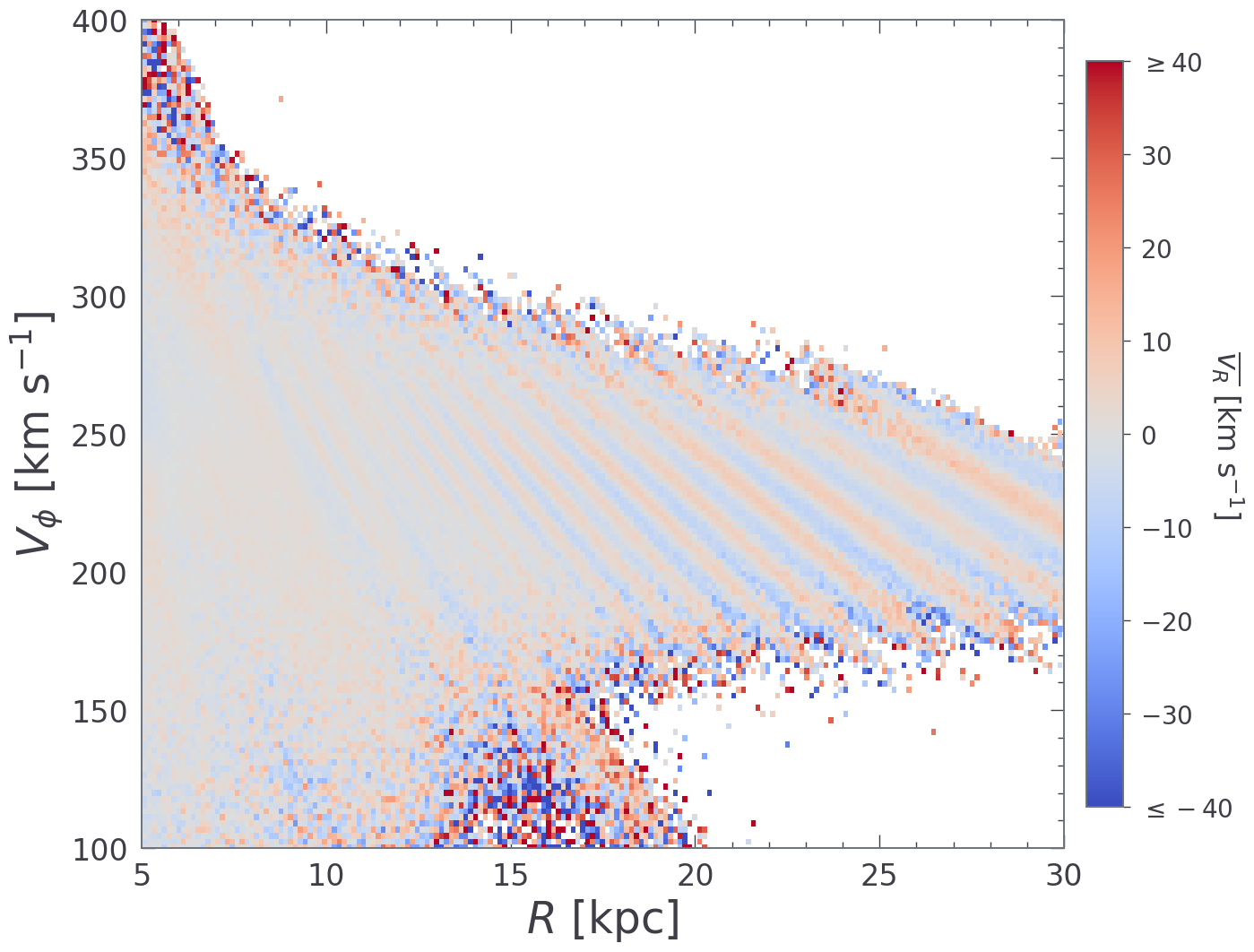}
 \centering
 \caption{Velocity waves in the $R-V_\phi$ plane colour-coded by $\overline{V}_R$ using the N-body simulations. The snapshot at $T=4.4\Gyr$ is selected. Note that no selection criteria are applied and all the disk particles are used in this plot.}
 \label{fig:nbody}
\end{figure}

\begin{figure}
 \includegraphics[width=\columnwidth]{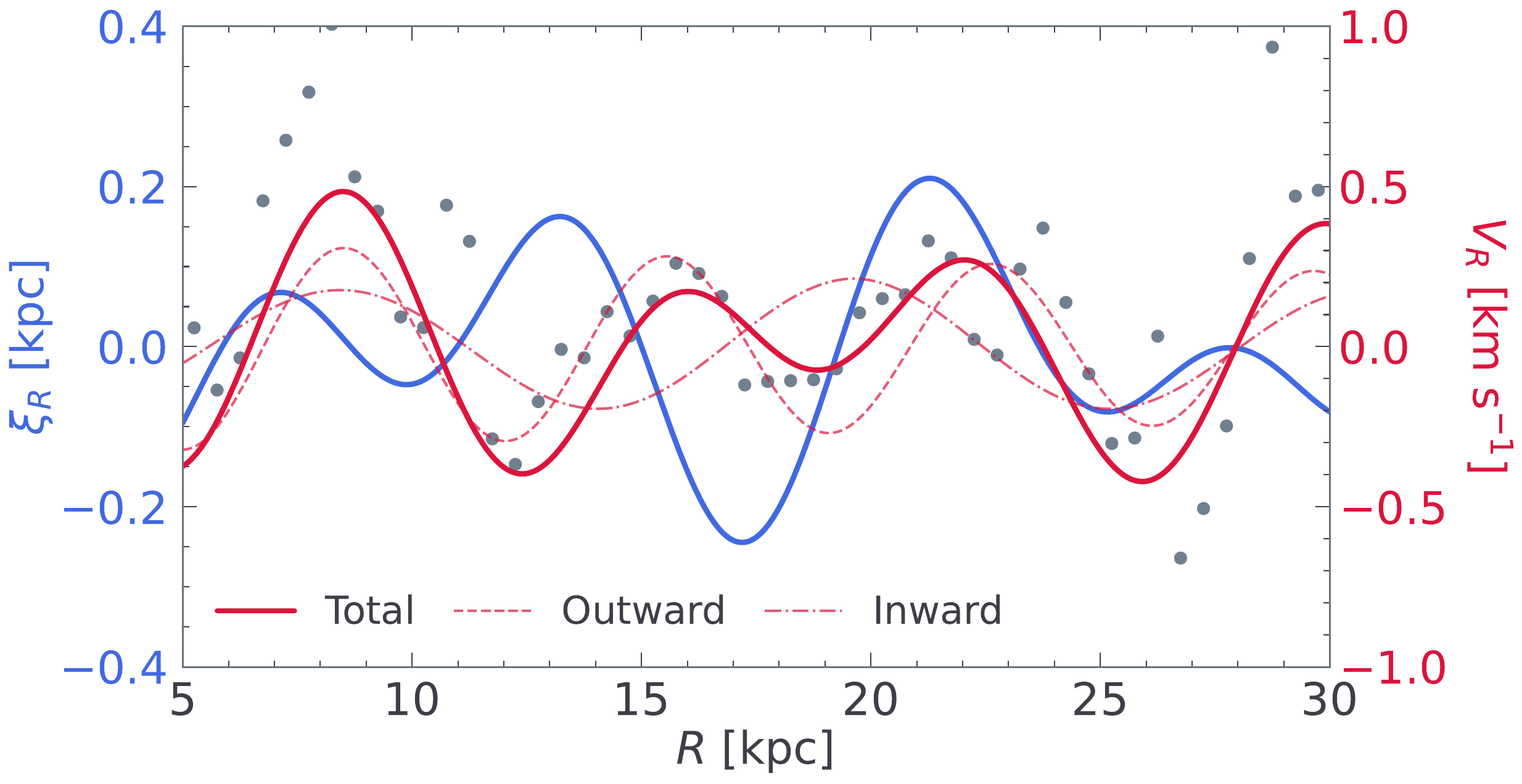}
 \centering
 \caption{Corrugation model fitted to the simulated sample at $T=4.4\Gyr$. The grey points show the median radial velocity in each radial bin. The red curve represents the best-fitting $V_{\mathrm{R}}$ pattern, and the blue curve shows the corresponding inferred radial displacement $\xi_R$, both plotted as functions of the galactocentric radius $R$.}
 \label{fig:radial_corrugation_sim}
\end{figure}

We successfully reproduced radial velocity oscillation as a function of galactocentric radius in the N-body simulation. 
Fig.~\ref{fig:nbody} shows the velocity wave in the $R-V_\phi$ plane colour-coded by $V_R$ using the N-body simulations. We chose the snapshot at $T=4.4\Gyr$  both in this plot and in Fig.~\ref{fig:radial_corrugation_sim}. The wave-like pattern is clearly seen both in the inner and outer disk. This snapshot does not represent the present-day properties of the MW disk, as both the amplitude and wavelength differ from those inferred from the observations. We chose this snapshot because it exhibits a clear large-scale kinematic pattern that is qualitatively consistent with the superposition of two oppositely propagating wave components, providing a useful phenomenological illustration for the interpretation of the observational results. As shown in Fig.~\ref{fig:radial_corrugation_sim}, the simulated sample exhibits a clear periodic variation with $R$, closely resembling the behaviour identified in the Gaia data. Following the same procedure adopted for the observational sample, we directly fitted the simulated $V_{\mathrm{R}}$ distribution with the model in Eq.~\ref{eq:vR_model_damped}.

The best-fitting parameters are listed in Table~\ref{tab:parameters_sim}. The amplitudes recovered from the simulation are slightly smaller than those inferred from the observations. The simulation also tends to favour larger damping lengths and smaller propagation velocities, although these parameters remain only weakly constrained in the toy-model fit. These discrepancies may reflect the simplified nature of the model as well as the differences between the simulated MW-Sgr interaction and the real Galaxy.

Importantly, despite these quantitative differences, both the simulations and the observations display comparable periodic variation in galactocentric radius and radial velocity, and the overall pattern of oscillation is broadly similar. As discussed above, our aim here is not to establish a detailed quantitative model, but rather to provide an intuitive dynamical demonstration of the plausibility of such a corrugation pattern. In this sense, the fitted parameters should be regarded primarily as an effective characterization of the overall wave pattern, rather than as precise physical measurements. Taken together, these results suggest that the superposition of an inward- and an outward-propagating wave, as encoded in our corrugation model, can provide a plausible phenomenological description of the transition identified in Sect.~\ref{sec:transition}.

\begin{table*}
\begin{center}
    \caption{Best-fitting parameters of the inward and outward corrugation modes in simulation.}
    \begin{tabular*}{\textwidth}{@{\extracolsep{\fill}}lccccc}
        \hline
        Mode & $A_{C}$ (kpc) & $\lambda_{C}$ (kpc) & $R_0$ (kpc) & $L_{C}$ (kpc) & $v_{C}$ (kpc Gyr$^{-1}$) \\
        \hline
        Inwards  & 0.13 & 11.3 & 19.6 & 60.0 & -3.0 \\
        Outwards & 0.15 & 7.1 & 5.0 & 80.0 & 2.5  \\
        \hline
    \end{tabular*}
\label{tab:parameters_sim}
\end{center}
\end{table*}

\section{Conclusion}\label{sec:conclusion}
In this work, we carried out a comprehensive investigation of the dual structure of the Galactic thin disk using large stellar samples from LAMOST DR8 and Gaia DR3. By jointly analysing spatial, kinematic, and chemical information, we identified a clear transition between the inner and outer disks. First, we reproduced the wave-like pattern in the $R_\mathrm{g}-V_\phi$ plane colour-coded by $V_\mathrm{R}$ reported by \citet{chen2024inward} and found a consistent transition at $R_\mathrm{g} \sim 13.5$ kpc in both datasets. This boundary is also imprinted in the $V_\phi-R_\mathrm{g}$ relation, where the trend of the rotational velocity changes near $R_\mathrm{g}\sim13.5$ kpc. Furthermore, Fig.~\ref{fig:vPhi_metallicity} shows that this transition is accompanied by a change in metallicity, indicating that it is not purely kinematic. Taken together, these observational signatures delineate a robust boundary separating the disk into two distinct kinematic regimes, corresponding to the inner and outer disk.

To provide physical insight, we constructed a simplified corrugation model and performed N-body simulations. Building upon the framework of \citet{poggio2025great}, our model incorporates two waves propagating in opposite directions. This formulation can be fitted to the observational data and applies equally well to the simulated sample. Both the observed and simulated disks exhibit coherent periodic variations in radial velocity and galactocentric radius, suggesting that the identified transition may be understood as part of a large-scale dynamical response of the disk. As discussed in Sect.~\ref{sec:link to corrugation}, such a corrugation could naturally produce a transition between inner and outer disk behaviours. Although not intended as a quantitative model, it provides an intuitive demonstration and offers a plausible interpretation of the observed features.

Overall, our results highlight the complex nature of the MW disk and reinforce the emerging view that its large-scale kinematic substructures are shaped by multiple time-dependent perturbations. The Galactic bar or possible external influences may contribute to the observed features. Future high-resolution spectroscopic surveys—particularly DESI and PFS—will provide crucial constraints on the chemical and kinematic signatures across the transition radius, enabling a more complete dynamical understanding of the Galactic disk.

\begin{acknowledgements}
This study is supported by the National Natural Science Foundation of China under grant Nos. 12588202 and 12373020, National Key Research and Development Program of China under grant Nos. 2024YFA1611902 and 2023YFE0107800. CL is supported by the Fundamental Research Funds for the Central Universities under grant No. KG202502. CY acknowledges the support from the National Natural Science Foundation of China under grant No. 12403025.

We have made use of data from the European Space Agency (ESA) mission
{\it Gaia} (\url{https://www.cosmos.esa.int/gaia}), processed by the {\it Gaia}
Data Processing and Analysis Consortium (DPAC,
\url{https://www.cosmos.esa.int/web/gaia/dpac/consortium}). Funding for the DPAC
has been provided by national institutions, in particular the institutions
participating in the {\it Gaia} Multilateral Agreement.
\par
This work also made use of data products from the Guo Shoujing Telescope (the Large Sky Area Multi-Object Fiber Spectroscopic Telescope, LAMOST). LAMOST is a National Major Scientific Project built by the Chinese Academy of Sciences. Funding for the project has been provided by the National Development and Reform Commission. LAMOST is operated and managed by the National Astronomical Observatories, Chinese Academy of Sciences.
\end{acknowledgements}

\bibliographystyle{aa}
\bibliography{references}

@article{BlandHawthorn2016,
  author = {{Bland-Hawthorn}, J. and {Gerhard}, O.},
  title   = {The galaxy in context: structural, kinematic, and integrated properties},
  journal = {Annual Review of Astronomy and Astrophysics},
  volume  = {54},
  number  = {1},
  pages   = {529--596},
  year    = {2016}
}

@article{chiappini1997chemical,
  title={The chemical evolution of the galaxy: the two-infall model},
  author={Chiappini, C. and Gratton, R.},
  journal={The Astrophysical Journal},
  volume={477},
  number={2},
  pages={765},
  year={1997},
  publisher={IOP Publishing}
}

@article{chiappini2001abundance,
  title={Abundance gradients and the formation of the Milky Way},
  author={Chiappini, Cristina and Matteucci, Francesca and Romano, Donatella},
  journal={The Astrophysical Journal},
  volume={554},
  number={2},
  pages={1044},
  year={2001},
  publisher={IOP Publishing}
}

@article{binney1998,
  title={Galactic astronomy},
  author={Binney, James and Merrifield, Michael},
  year={1998},
  publisher={Princeton University Press}
}

@article{katz2021radial,
  title={Radial structure and formation of the Milky Way disc},
  author={Katz, D and G{\'o}mez, A and Haywood, M and Snaith, O and Di Matteo, P},
  journal={Astronomy \& Astrophysics},
  volume={655},
  pages={A111},
  year={2021},
  publisher={EDP Sciences}
}

@article{Majewski2017,
  author = {{Majewski}, S. R. and {Schiavon}, R. P. and {Frinchaboy}, P. M. and et {al.}},
  title = {The apache point observatory galactic evolution experiment (APOGEE)},
  journal = {The Astronomical Journal},
  volume = {154},
  number = {3},
  pages = {94},
  year = {2017}
}

@article{huang2015metallicity,
  title={On the metallicity gradients of the Galactic disk as revealed by LSS-GAC red clump stars},
  author={Huang, Yang and Liu, Xiao-Wei and Zhang, Hua-Wei and Yuan, Hai-Bo and Xiang, Mao-Sheng and Chen, Bing-Qiu and Ren, Juan-Juan and Sun, Ning-Chen and Wang, Chun and Zhang, Yong and others},
  journal={Research in Astronomy and Astrophysics},
  volume={15},
  number={8},
  pages={1240},
  year={2015},
  publisher={IOP Publishing}
}

@article{chen2024inward,
  title={An Inward-moving and Asymmetric Velocity Wave Detected in LAMOST-Gaia},
  author={Chen, Yuqin and Zhao, Gang and Wu, Wenbo and Guo, Zixi and Zhang, Haopeng and Li, Zhuohan},
  journal={The Astrophysical Journal Letters},
  volume={961},
  number={1},
  pages={L4},
  year={2024},
  publisher={IOP Publishing}
}

@article{zhang2021radial,
  title={Radial migration from the metallicity gradient of open clusters and outliers},
  author={Zhang, Haopeng and Chen, Yuqin and Zhao, Gang},
  journal={The Astrophysical Journal},
  volume={919},
  number={1},
  pages={52},
  year={2021},
  publisher={IOP Publishing}
}

@article{frankel2019inside,
  title={The inside-out growth of the Galactic disk},
  author={Frankel, Neige and Sanders, Jason and Rix, Hans-Walter and Ting, Yuan-Sen and Ness, Melissa},
  journal={The Astrophysical Journal},
  volume={884},
  number={2},
  pages={99},
  year={2019},
  publisher={IOP Publishing}
}

@article{spina2022mapping,
  title={Mapping the galactic metallicity gradient with open clusters: the state-of-the-art and future challenges},
  author={Spina, Lorenzo and Magrini, Laura and Cunha, Katia},
  journal={Universe},
  volume={8},
  number={2},
  pages={87},
  year={2022},
  publisher={MDPI}
}

@article{das2024outer,
  title={The outer low-$\alpha$ disc of the Milky Way--I: evidence for the first pericentric passage of Sagittarius?},
  author={Das, Payel and Huang, Yang and Ciuc{\u{a}}, Ioana and Fragkoudi, Francesca},
  journal={Monthly Notices of the Royal Astronomical Society},
  volume={527},
  number={3},
  pages={4505--4514},
  year={2024},
  publisher={Oxford University Press}
}

@article{reid2019trigonometric,
  title={Trigonometric parallaxes of high-mass star-forming regions: our view of the Milky Way},
  author={Reid, MJ and Menten, KM and Brunthaler, A and Zheng, XW and Dame, TM and Xu, Y and Li, J and Sakai, N and Wu, Y and Immer, K and others},
  journal={The Astrophysical Journal},
  volume={885},
  number={2},
  pages={131},
  year={2019},
  publisher={IOP Publishing}
}

@article{poggio2025great,
  title={The great wave-Evidence of a large-scale vertical corrugation propagating outwards in the Galactic disc},
  author={Poggio, E and Khanna, S and Drimmel, R and Zari, E and D’Onghia, E and Lattanzi, MG and Palicio, PA and Recio-Blanco, A and Thulasidharan, L},
  journal={Astronomy \& Astrophysics},
  volume={699},
  pages={A199},
  year={2025},
  publisher={EDP Sciences}
}

@article{li2022stellar,
  title={The stellar parameters and elemental abundances from low-resolution spectra--I. 1.2 million giants from LAMOST DR8},
  author={Li, Zhuohan and Zhao, Gang and Chen, Yuqin and Liang, Xilong and Zhao, Jingkun},
  journal={Monthly Notices of the Royal Astronomical Society},
  volume={517},
  number={4},
  pages={4875--4891},
  year={2022},
  publisher={Oxford University Press}
}

@article{li2024exploring,
  title={Exploring the ex-situ components within Gaia DR3},
  author={Li, Zhuohan and Zhao, Gang and Zhang, Ruizhi and Xue, Xiang-Xiang and Chen, Yuqin and Amarante, Jo{\~a}o AS},
  journal={Monthly Notices of the Royal Astronomical Society},
  volume={527},
  number={4},
  pages={9767--9781},
  year={2024},
  publisher={Oxford University Press}
}

@article{lindegren2021gaia,
  title={Gaia Early Data Release 3-Parallax bias versus magnitude, colour, and position},
  author={Lindegren, L and Bastian, U and Biermann, M and Bombrun, A and De Torres, A and Gerlach, E and Geyer, R and Hern{\'a}ndez, J and Hilger, T and Hobbs, D and others},
  journal={Astronomy \& Astrophysics},
  volume={649},
  pages={A4},
  year={2021},
  publisher={EDP Sciences}
}

@ARTICLE{TheGaiamission,
       author = {{Gaia Collaboration} and others},
        title = "{The Gaia mission}",
      journal = {\aap},
         year = 2016,
        month = nov,
       volume = {595},
          eid = {A1},
        pages = {A1},
          doi = {10.1051/0004-6361/201629272},
archivePrefix = {arXiv},
       eprint = {1609.04153},
 primaryClass = {astro-ph.IM},
       adsurl = {https://ui.adsabs.harvard.edu/abs/2016A&A...595A...1G}
}

@ARTICLE{GaiaDR3,
       author = {{Gaia Collaboration} and others},
        title = "{Gaia Data Release 3. Summary of the content and survey properties}",
      journal = {\aap},
         year = 2023,
        month = jun,
       volume = {674},
          eid = {A1},
        pages = {A1},
          doi = {10.1051/0004-6361/202243940},
archivePrefix = {arXiv},
       eprint = {2208.00211},
 primaryClass = {astro-ph.GA},
       adsurl = {https://ui.adsabs.harvard.edu/abs/2023A&A...674A...1G}
}

@article{luo2015first,
  title={The first data release (DR1) of the LAMOST regular survey},
  author={Luo, A-Li and Zhao, Yong-Heng and Zhao, Gang and Deng, Li-Cai and Liu, Xiao-Wei and Jing, Yi-Peng and Wang, Gang and Zhang, Hao-Tong and Shi, Jian-Rong and Cui, Xiang-Qun and others},
  journal={Research in Astronomy and Astrophysics},
  volume={15},
  number={8},
  pages={1095},
  year={2015},
  publisher={IOP Publishing}
}

@article{zhao2012lamost,
  title={LAMOST spectral survey—An overview},
  author={Zhao, Gang and Zhao, Yong-Heng and Chu, Yao-Quan and Jing, Yi-Peng and Deng, Li-Cai},
  journal={Research in Astronomy and Astrophysics},
  volume={12},
  number={7},
  pages={723},
  year={2012},
  publisher={IOP Publishing}
}

@article{anders2022photo,
  title={Photo-astrometric distances, extinctions, and astrophysical parameters for Gaia EDR3 stars brighter than G= 18.5},
  author={Anders, Friedrich and Khalatyan, A and Queiroz, Anna B{\'a}rbara de Andrade and Chiappini, Cristina and Ard{\`e}vol, J and Casamiquela, Laia and Figueras, F and Jim{\'e}nez-Arranz, {\'O} and Jordi, Carme and Mongui{\'o}, M and others},
  journal={Astronomy \& Astrophysics},
  volume={658},
  pages={A91},
  year={2022},
  publisher={EDP Sciences}
}

@ARTICLE{StarHorse,
       author = {{Queiroz}, A.~B.~A. and {Anders}, F. and {Chiappini}, C. and {Khalatyan}, A. and {Santiago}, B.~X. and {Steinmetz}, M. and {Valentini}, M. and {Miglio}, A. and {Bossini}, D. and {Barbuy}, B. and {Minchev}, I. and {Minniti}, D. and {Garc{\'\i}a Hern{\'a}ndez}, D.~A. and {Schultheis}, M. and {Beaton}, R.~L. and {Beers}, T.~C. and {Bizyaev}, D. and {Brownstein}, J.~R. and {Cunha}, K. and {Fern{\'a}ndez-Trincado}, J.~G. and {Frinchaboy}, P.~M. and {Lane}, R.~R. and {Majewski}, S.~R. and {Nataf}, D. and {Nitschelm}, C. and {Pan}, K. and {Roman-Lopes}, A. and {Sobeck}, J.~S. and {Stringfellow}, G. and {Zamora}, O.},
        title = "{From the bulge to the outer disc: StarHorse stellar parameters, distances, and extinctions for stars in APOGEE DR16 and other spectroscopic surveys}",
      journal = {\aap},
         year = 2020,
        month = jun,
       volume = {638},
          eid = {A76},
        pages = {A76},
          doi = {10.1051/0004-6361/201937364},
archivePrefix = {arXiv},
       eprint = {1912.09778},
 primaryClass = {astro-ph.GA},
       adsurl = {https://ui.adsabs.harvard.edu/abs/2020A&A...638A..76Q}
}

@article{accetta2022seventeenth,
  title={The seventeenth data release of the Sloan Digital Sky Surveys: Complete release of MaNGA, MaStar, and APOGEE-2 data},
  author={Accetta, Katherine and Aerts, Conny and Aguirre, Victor Silva and Ahumada, Romina and Ajgaonkar, Nikhil and Ak, N Filiz and Alam, Shadab and Prieto, Carlos Allende and Almeida, Andres and Anders, Friedrich and others},
  journal={The Astrophysical Journal Supplement Series},
  volume={259},
  number={2},
  pages={35},
  year={2022},
  publisher={IOP Publishing}
}

@ARTICLE{Li2025,
       author = {{Li}, Zhuohan and {Li}, Chengdong and {Zhao}, Gang and {Zhang}, Ruizhi and {Xue}, Xiang-Xiang},
        title = "{The Rotating Bulge and Halo in the Milky Way: Evidence of Angular Momentum Transferred from the Decelerating Bar}",
      journal = {\apj},
         year = 2025,
        month = jul,
       volume = {988},
       number = {1},
          eid = {124},
        pages = {124},
          doi = {10.3847/1538-4357/ade430},
archivePrefix = {arXiv},
       eprint = {2506.12717},
 primaryClass = {astro-ph.GA},
       adsurl = {https://ui.adsabs.harvard.edu/abs/2025ApJ...988..124L}
}

@article{mcmillan2016mass,
  title={The mass distribution and gravitational potential of the Milky Way},
  author={McMillan, Paul J},
  journal={Monthly Notices of the Royal Astronomical Society},
  volume={465},
  pages={76--94},
  year={2017},
  publisher={Oxford University Press}
}

@ARTICLE{Bennett19,
       author = {{Bennett}, Morgan and {Bovy}, Jo},
        title = "{Vertical waves in the solar neighbourhood in Gaia DR2}",
      journal = {\mnras},
         year = 2019,
        month = jan,
       volume = {482},
       number = {1},
        pages = {1417-1425},
          doi = {10.1093/mnras/sty2813},
archivePrefix = {arXiv},
       eprint = {1809.03507},
 primaryClass = {astro-ph.GA},
       adsurl = {https://ui.adsabs.harvard.edu/abs/2019MNRAS.482.1417B}
}

@article{schonrich2010local,
  title={Local kinematics and the local standard of rest},
  author={Sch{\"o}nrich, Ralph and Binney, James and Dehnen, Walter},
  journal={Monthly Notices of the Royal Astronomical Society},
  volume={403},
  number={4},
  pages={1829--1833},
  year={2010},
  publisher={The Royal Astronomical Society}
}

@article{andrae2023gaia,
  title={Gaia data release 3-analysis of the gaia bp/rp spectra using the general stellar parameterizer from photometry},
  author={Andrae, Ren{\'e} and Fouesneau, Morgan and Sordo, Rosanna and Bailer-Jones, Coryn AL and Dharmawardena, TE and Rybizki, J and De Angeli, Francesca and Lindstr{\o}m, HEP and Marshall, Douglas J and Drimmel, R and others},
  journal={Astronomy \& Astrophysics},
  volume={674},
  pages={A27},
  year={2023},
  publisher={EDP Sciences}
}

@ARTICLE{Binney2023,
       author = {{Binney}, James and {Vasiliev}, Eugene},
        title = "{Self-consistent models of our Galaxy}",
      journal = {\mnras},
         year = 2023,
        month = apr,
       volume = {520},
       number = {2},
        pages = {1832-1847},
          doi = {10.1093/mnras/stad094},
archivePrefix = {arXiv},
       eprint = {2206.03523},
 primaryClass = {astro-ph.GA},
       adsurl = {https://ui.adsabs.harvard.edu/abs/2023MNRAS.520.1832B}
}

@ARTICLE{Binney2024,
       author = {{Binney}, James and {Vasiliev}, Eugene},
        title = "{Chemodynamical models of our Galaxy}",
      journal = {\mnras},
         year = 2024,
        month = jan,
       volume = {527},
       number = {2},
        pages = {1915-1934},
          doi = {10.1093/mnras/stad3312},
archivePrefix = {arXiv},
       eprint = {2306.11602},
       adsurl = {https://ui.adsabs.harvard.edu/abs/2024MNRAS.527.1915B}
}

@BOOK{Binney2008,
       author = {{Binney}, James and {Tremaine}, Scott},
        title = "{Galactic Dynamics: Second Edition}",
         year = 2008,
       adsurl = {https://ui.adsabs.harvard.edu/abs/2008gady.book.....B}
}

@ARTICLE{Vasiliev2019,
       author = {{Vasiliev}, Eugene},
        title = "{AGAMA: action-based galaxy modelling architecture}",
      journal = {\mnras},
         year = 2019,
        month = jan,
       volume = {482},
       number = {2},
        pages = {1525-1544},
          doi = {10.1093/mnras/sty2672},
archivePrefix = {arXiv},
       eprint = {1802.08239},
 primaryClass = {astro-ph.GA},
       adsurl = {https://ui.adsabs.harvard.edu/abs/2019MNRAS.482.1525V}
}

@ARTICLE{Li2022a,
       author = {{Li}, Chengdong and {Binney}, James},
        title = "{Modelling the stellar halo with RR-Lyrae stars}",
      journal = {\mnras},
         year = 2022,
        month = mar,
       volume = {510},
       number = {4},
        pages = {4706-4722},
          doi = {10.1093/mnras/stab3711},
archivePrefix = {arXiv},
       eprint = {2109.02324},
 primaryClass = {astro-ph.GA},
       adsurl = {https://ui.adsabs.harvard.edu/abs/2022MNRAS.510.4706L}
}

@ARTICLE{Li2022b,
       author = {{Li}, Chengdong and {Binney}, James},
        title = "{Our Galaxy's youngest disc}",
      journal = {\mnras},
         year = 2022,
        month = nov,
       volume = {516},
       number = {3},
        pages = {3454-3469},
          doi = {10.1093/mnras/stac1788},
archivePrefix = {arXiv},
       eprint = {2205.01455},
 primaryClass = {astro-ph.GA},
       adsurl = {https://ui.adsabs.harvard.edu/abs/2022MNRAS.516.3454L}
}

@ARTICLE{Ablimit2020,
       author = {{Ablimit}, Iminhaji and {Zhao}, Gang and {Flynn}, Chris and {Bird}, Sarah A.},
        title = "{The Rotation Curve, Mass Distribution, and Dark Matter Content of the Milky Way from Classical Cepheids}",
      journal = {\apjl},
         year = 2020,
        month = may,
       volume = {895},
       number = {1},
          eid = {L12},
        pages = {L12},
          doi = {10.3847/2041-8213/ab8d45},
archivePrefix = {arXiv},
       eprint = {2004.13768},
 primaryClass = {astro-ph.GA},
       adsurl = {https://ui.adsabs.harvard.edu/abs/2020ApJ...895L..12A}
}

@ARTICLE{Eilers2019,
       author = {{Eilers}, Anna-Christina and {Hogg}, David W. and {Rix}, Hans-Walter and {Ness}, Melissa K.},
        title = "{The Circular Velocity Curve of the Milky Way from 5 to 25 kpc}",
      journal = {\apj},
         year = 2019,
        month = jan,
       volume = {871},
       number = {1},
          eid = {120},
        pages = {120},
          doi = {10.3847/1538-4357/aaf648},
archivePrefix = {arXiv},
       eprint = {1810.09466},
 primaryClass = {astro-ph.GA},
       adsurl = {https://ui.adsabs.harvard.edu/abs/2019ApJ...871..120E}
}

@ARTICLE{Springel2021,
       author = {{Springel}, Volker and {Pakmor}, R{\"u}diger and {Zier}, Oliver and {Reinecke}, Martin},
        title = "{Simulating cosmic structure formation with the GADGET-4 code}",
      journal = {\mnras},
         year = 2021,
        month = sep,
       volume = {506},
       number = {2},
        pages = {2871-2949},
          doi = {10.1093/mnras/stab1855},
archivePrefix = {arXiv},
       eprint = {2010.03567},
 primaryClass = {astro-ph.IM},
       adsurl = {https://ui.adsabs.harvard.edu/abs/2021MNRAS.506.2871S}
}

@ARTICLE{Moster14,
       author = {{Moster}, Benjamin P. and {Macci{\`o}}, Andrea V. and {Somerville}, Rachel S.},
        title = "{Numerical hydrodynamic simulations based on semi-analytic galaxy merger trees: method and Milky Way-like galaxies}",
      journal = {\mnras},
         year = 2014,
        month = jan,
       volume = {437},
       number = {2},
        pages = {1027-1044},
          doi = {10.1093/mnras/stt1702},
archivePrefix = {arXiv},
       eprint = {1210.5522},
 primaryClass = {astro-ph.CO},
       adsurl = {https://ui.adsabs.harvard.edu/abs/2014MNRAS.437.1027M}
}

@ARTICLE{Binney2014,
       author = {{Binney}, James},
        title = "{Self-consistent flattened isochrones}",
      journal = {\mnras},
         year = 2014,
        month = may,
       volume = {440},
       number = {1},
        pages = {787-798},
          doi = {10.1093/mnras/stu297},
archivePrefix = {arXiv},
       eprint = {1402.2512},
 primaryClass = {astro-ph.GA},
       adsurl = {https://ui.adsabs.harvard.edu/abs/2014MNRAS.440..787B}
}

@ARTICLE{Vasiliev21,
       author = {{Vasiliev}, Eugene and {Belokurov}, Vasily and {Erkal}, Denis},
        title = "{Tango for three: Sagittarius, LMC, and the Milky Way}",
      journal = {\mnras},
         year = 2021,
        month = feb,
       volume = {501},
       number = {2},
        pages = {2279-2304},
          doi = {10.1093/mnras/staa3673},
archivePrefix = {arXiv},
       eprint = {2009.10726},
 primaryClass = {astro-ph.GA},
       adsurl = {https://ui.adsabs.harvard.edu/abs/2021MNRAS.501.2279V}
}

@ARTICLE{Chandrasekhar1943,
       author = {{Chandrasekhar}, S.},
        title = "{Dynamical Friction. I. General Considerations: the Coefficient of Dynamical Friction.}",
      journal = {\apj},
         year = 1943,
        month = mar,
       volume = {97},
        pages = {255},
          doi = {10.1086/144517},
       adsurl = {https://ui.adsabs.harvard.edu/abs/1943ApJ....97..255C}
}

@article{tian2024mapping,
  title={Mapping the Milky Way with LAMOST--IV. The large Galactic disc extending to 35 kpc},
  author={Tian, Hao and Liu, Chao and Li, Jiadong and Zhang, Bo},
  journal={Monthly Notices of the Royal Astronomical Society},
  volume={531},
  number={1},
  pages={1730--1745},
  year={2024},
  publisher={Oxford University Press}
}

@ARTICLE{JT1966,
       author = {{Julian}, William H. and {Toomre}, Alar},
        title = "{Non-Axisymmetric Responses of Differentially Rotating Disks of Stars}",
      journal = {\apj},
         year = 1966,
        month = dec,
       volume = {146},
        pages = {810},
          doi = {10.1086/148957},
       adsurl = {https://ui.adsabs.harvard.edu/abs/1966ApJ...146..810J}
}

@ARTICLE{LS1966,
       author = {{Lin}, C.~C. and {Shu}, Frank H.},
        title = "{On the Spiral Structure of Disk Galaxies, II. Outline of a Theory of Density Waves}",
      journal = {Proceedings of the National Academy of Science},
         year = 1966,
        month = feb,
       volume = {55},
       number = {2},
        pages = {229-234},
          doi = {10.1073/pnas.55.2.229},
       adsurl = {https://ui.adsabs.harvard.edu/abs/1966PNAS...55..229L}
}

@ARTICLE{Toomre1969,
       author = {{Toomre}, Alar},
        title = "{Group Velocity of Spiral Waves in Galactic Disks}",
      journal = {\apj},
         year = 1969,
        month = dec,
       volume = {158},
        pages = {899},
          doi = {10.1086/150250},
       adsurl = {https://ui.adsabs.harvard.edu/abs/1969ApJ...158..899T}
}

@ARTICLE{LBK1972,
       author = {{Lynden-Bell}, D. and {Kalnajs}, A.~J.},
        title = "{On the generating mechanism of spiral structure}",
      journal = {\mnras},
         year = 1972,
        month = jan,
       volume = {157},
        pages = {1},
          doi = {10.1093/mnras/157.1.1},
       adsurl = {https://ui.adsabs.harvard.edu/abs/1972MNRAS.157....1L}
}

@ARTICLE{Toomre1977,
       author = {{Toomre}, A.},
        title = "{Theories of spiral structure.}",
      journal = {\araa},
         year = 1977,
        month = jan,
       volume = {15},
        pages = {437-478},
          doi = {10.1146/annurev.aa.15.090177.002253},
       adsurl = {https://ui.adsabs.harvard.edu/abs/1977ARA&A..15..437T}
}

@INPROCEEDINGS{Toomre1981,
       author = {{Toomre}, A.},
        title = "{What amplifies the spirals}",
    booktitle = {Structure and Evolution of Normal Galaxies},
         year = 1981,
       editor = {{Fall}, S.~M. and {Lynden-Bell}, D.},
        month = jan,
        pages = {111-136},
       adsurl = {https://ui.adsabs.harvard.edu/abs/1981seng.proc..111T}
}

@ARTICLE{Antoja2018,
       author = {{Antoja}, T. and {Helmi}, A. and {Romero-G{\'o}mez}, M. and {Katz}, D. and {Babusiaux}, C. and {Drimmel}, R. and {Evans}, D.~W. and {Figueras}, F. and {Poggio}, E. and {Reyl{\'e}}, C. and {Robin}, A.~C. and {Seabroke}, G. and {Soubiran}, C.},
        title = "{A dynamically young and perturbed Milky Way disk}",
      journal = {\nat},
         year = 2018,
        month = sep,
       volume = {561},
       number = {7723},
        pages = {360-362},
          doi = {10.1038/s41586-018-0510-7},
archivePrefix = {arXiv},
       eprint = {1804.10196},
 primaryClass = {astro-ph.GA},
       adsurl = {https://ui.adsabs.harvard.edu/abs/2018Natur.561..360A}
}

@ARTICLE{Binney2018,
       author = {{Binney}, James and {Sch{\"o}nrich}, Ralph},
        title = "{The origin of the Gaia phase-plane spiral}",
      journal = {\mnras},
         year = 2018,
        month = dec,
       volume = {481},
       number = {2},
        pages = {1501-1506},
          doi = {10.1093/mnras/sty2378},
archivePrefix = {arXiv},
       eprint = {1807.09819},
 primaryClass = {astro-ph.GA},
       adsurl = {https://ui.adsabs.harvard.edu/abs/2018MNRAS.481.1501B}
}

@ARTICLE{Hunt2022,
       author = {{Hunt}, Jason A.~S. and {Price-Whelan}, Adrian M. and {Johnston}, Kathryn V. and {Darragh-Ford}, Elise},
        title = "{Multiple phase spirals suggest multiple origins in Gaia DR3}",
      journal = {\mnras},
         year = 2022,
        month = oct,
       volume = {516},
       number = {1},
        pages = {L7-L11},
          doi = {10.1093/mnrasl/slac082},
archivePrefix = {arXiv},
       eprint = {2206.06125},
 primaryClass = {astro-ph.GA},
       adsurl = {https://ui.adsabs.harvard.edu/abs/2022MNRAS.516L...7H}
}

@ARTICLE{Khanna2019,
       author = {{Khanna}, Shourya and {Sharma}, Sanjib and {Tepper-Garcia}, Thor and {Bland-Hawthorn}, Joss and {Hayden}, Michael and {Asplund}, Martin and {Buder}, Sven and {Chen}, Boquan and {De Silva}, Gayandhi M. and {Freeman}, Ken C. and {Kos}, Janez and {Lewis}, Geraint F. and {Lin}, Jane and {Martell}, Sarah L. and {Simpson}, Jeffrey D. and {Nordlander}, Thomas and {Stello}, Dennis and {Ting}, Yuan-Sen and {Zucker}, Daniel B. and {Zwitter}, Toma{\v{z}}},
        title = "{The GALAH survey and Gaia DR2: Linking ridges, arches, and vertical waves in the kinematics of the Milky Way}",
      journal = {\mnras},
         year = 2019,
        month = nov,
       volume = {489},
       number = {4},
        pages = {4962-4979},
          doi = {10.1093/mnras/stz2462},
archivePrefix = {arXiv},
       eprint = {1902.10113},
 primaryClass = {astro-ph.GA},
       adsurl = {https://ui.adsabs.harvard.edu/abs/2019MNRAS.489.4962K}
}

@ARTICLE{Laporte2019,
       author = {{Laporte}, Chervin F.~P. and {Minchev}, Ivan and {Johnston}, Kathryn V. and {G{\'o}mez}, Facundo A.},
        title = "{Footprints of the Sagittarius dwarf galaxy in the Gaia data set}",
      journal = {\mnras},
         year = 2019,
        month = may,
       volume = {485},
       number = {3},
        pages = {3134-3152},
          doi = {10.1093/mnras/stz583},
archivePrefix = {arXiv},
       eprint = {1808.00451},
 primaryClass = {astro-ph.GA},
       adsurl = {https://ui.adsabs.harvard.edu/abs/2019MNRAS.485.3134L}
}

@ARTICLE{BlandHawthorn2021,
       author = {{Bland-Hawthorn}, Joss and {Tepper-Garc{\'\i}a}, Thor},
        title = "{Galactic seismology: the evolving 'phase spiral' after the Sagittarius dwarf impact}",
      journal = {\mnras},
         year = 2021,
        month = jul,
       volume = {504},
       number = {3},
        pages = {3168-3186},
          doi = {10.1093/mnras/stab704},
archivePrefix = {arXiv},
       eprint = {2009.02434},
 primaryClass = {astro-ph.GA},
       adsurl = {https://ui.adsabs.harvard.edu/abs/2021MNRAS.504.3168B}
}

@ARTICLE{Banik2022,
       author = {{Banik}, Uddipan and {Weinberg}, Martin D. and {van den Bosch}, Frank C.},
        title = "{A Comprehensive Perturbative Formalism for Phase Mixing in Perturbed Disks. I. Phase Spirals in an Infinite, Isothermal Slab}",
      journal = {\apj},
         year = 2022,
        month = aug,
       volume = {935},
       number = {2},
          eid = {135},
        pages = {135},
          doi = {10.3847/1538-4357/ac7ff9},
archivePrefix = {arXiv},
       eprint = {2208.05038},
 primaryClass = {astro-ph.GA},
       adsurl = {https://ui.adsabs.harvard.edu/abs/2022ApJ...935..135B}
}

@ARTICLE{Li2023,
       author = {{Li}, Chengdong and {Siebert}, Arnaud and {Monari}, Giacomo and {Famaey}, Benoit and {Rozier}, Simon},
        title = "{Gaia DR3 features of the phase spiral and its possible relation to internal perturbations}",
      journal = {\mnras},
         year = 2023,
        month = oct,
       volume = {524},
       number = {4},
        pages = {6331-6344},
          doi = {10.1093/mnras/stad2199},
archivePrefix = {arXiv},
       eprint = {2303.06393},
 primaryClass = {astro-ph.GA},
       adsurl = {https://ui.adsabs.harvard.edu/abs/2023MNRAS.524.6331L}
}

@ARTICLE{Frankel2025,
       author = {{Frankel}, Neige and {Hogg}, David W. and {Tremaine}, Scott and {Price-Whelan}, Adrian and {Shen}, Jeff},
        title = "{Iron Snails: Nonequilibrium Dynamics and Spiral Abundance Patterns}",
      journal = {\apj},
         year = 2025,
        month = jul,
       volume = {987},
       number = {1},
          eid = {81},
        pages = {81},
          doi = {10.3847/1538-4357/add5ea},
archivePrefix = {arXiv},
       eprint = {2407.07149},
 primaryClass = {astro-ph.GA},
       adsurl = {https://ui.adsabs.harvard.edu/abs/2025ApJ...987...81F}
}

@ARTICLE{Chiba2025,
       author = {{Chiba}, Rimpei and {Frankel}, Neige and {Hamilton}, Chris},
        title = "{Origin of the two-armed vertical phase spiral in the inner Galactic disc}",
      journal = {\mnras},
         year = 2025,
        month = nov,
       volume = {543},
       number = {3},
        pages = {2159-2179},
          doi = {10.1093/mnras/staf1566},
archivePrefix = {arXiv},
       eprint = {2503.20869},
 primaryClass = {astro-ph.GA},
       adsurl = {https://ui.adsabs.harvard.edu/abs/2025MNRAS.543.2159C}
}

@ARTICLE{York2000,
       author = {{York}, Donald G. and {Adelman}, J. and {Anderson}, Jr., John E. and {Anderson}, Scott F. and {Annis}, James and {Bahcall}, Neta A. and {Bakken}, J.~A. and {Barkhouser}, Robert and {Bastian}, Steven and {Berman}, Eileen and {Boroski}, William N. and {Bracker}, Steve and {Briegel}, Charlie and {Briggs}, John W. and {Brinkmann}, J. and {Brunner}, Robert and {Burles}, Scott and {Carey}, Larry and {Carr}, Michael A. and {Castander}, Francisco J. and {Chen}, Bing and {Colestock}, Patrick L. and {Connolly}, A.~J. and {Crocker}, J.~H. and {Csabai}, Istv{\'a}n and {Czarapata}, Paul C. and {Davis}, John Eric and {Doi}, Mamoru and {Dombeck}, Tom and {Eisenstein}, Daniel and {Ellman}, Nancy and {Elms}, Brian R. and {Evans}, Michael L. and {Fan}, Xiaohui and {Federwitz}, Glenn R. and {Fiscelli}, Larry and {Friedman}, Scott and {Frieman}, Joshua A. and {Fukugita}, Masataka and {Gillespie}, Bruce and {Gunn}, James E. and {Gurbani}, Vijay K. and {de Haas}, Ernst and {Haldeman}, Merle and {Harris}, Frederick H. and {Hayes}, J. and {Heckman}, Timothy M. and {Hennessy}, G.~S. and {Hindsley}, Robert B. and {Holm}, Scott and {Holmgren}, Donald J. and {Huang}, Chi-hao and {Hull}, Charles and {Husby}, Don and {Ichikawa}, Shin-Ichi and {Ichikawa}, Takashi and {Ivezi{\'c}}, {\v{Z}}eljko and {Kent}, Stephen and {Kim}, Rita S.~J. and {Kinney}, E. and {Klaene}, Mark and {Kleinman}, A.~N. and {Kleinman}, S. and {Knapp}, G.~R. and {Korienek}, John and {Kron}, Richard G. and {Kunszt}, Peter Z. and {Lamb}, D.~Q. and {Lee}, B. and {Leger}, R. French and {Limmongkol}, Siriluk and {Lindenmeyer}, Carl and {Long}, Daniel C. and {Loomis}, Craig and {Loveday}, Jon and {Lucinio}, Rich and {Lupton}, Robert H. and {MacKinnon}, Bryan and {Mannery}, Edward J. and {Mantsch}, P.~M. and {Margon}, Bruce and {McGehee}, Peregrine and {McKay}, Timothy A. and {Meiksin}, Avery and {Merelli}, Aronne and {Monet}, David G. and {Munn}, Jeffrey A. and {Narayanan}, Vijay K. and {Nash}, Thomas and {Neilsen}, Eric and {Neswold}, Rich and {Newberg}, Heidi Jo and {Nichol}, R.~C. and {Nicinski}, Tom and {Nonino}, Mario and {Okada}, Norio and {Okamura}, Sadanori and {Ostriker}, Jeremiah P. and {Owen}, Russell and {Pauls}, A. George and {Peoples}, John and {Peterson}, R.~L. and {Petravick}, Donald and {Pier}, Jeffrey R. and {Pope}, Adrian and {Pordes}, Ruth and {Prosapio}, Angela and {Rechenmacher}, Ron and {Quinn}, Thomas R. and {Richards}, Gordon T. and {Richmond}, Michael W. and {Rivetta}, Claudio H. and {Rockosi}, Constance M. and {Ruthmansdorfer}, Kurt and {Sandford}, Dale and {Schlegel}, David J. and {Schneider}, Donald P. and {Sekiguchi}, Maki and {Sergey}, Gary and {Shimasaku}, Kazuhiro and {Siegmund}, Walter A. and {Smee}, Stephen and {Smith}, J. Allyn and {Snedden}, S. and {Stone}, R. and {Stoughton}, Chris and {Strauss}, Michael A. and {Stubbs}, Christopher and {SubbaRao}, Mark and {Szalay}, Alexander S. and {Szapudi}, Istvan and {Szokoly}, Gyula P. and {Thakar}, Anirudda R. and {Tremonti}, Christy and {Tucker}, Douglas L. and {Uomoto}, Alan and {Vanden Berk}, Dan and {Vogeley}, Michael S. and {Waddell}, Patrick and {Wang}, Shu-i. and {Watanabe}, Masaru and {Weinberg}, David H. and {Yanny}, Brian and {Yasuda}, Naoki and {SDSS Collaboration}},
        title = "{The Sloan Digital Sky Survey: Technical Summary}",
      journal = {\aj},
         year = 2000,
        month = sep,
       volume = {120},
       number = {3},
        pages = {1579-1587},
          doi = {10.1086/301513},
archivePrefix = {arXiv},
       eprint = {astro-ph/0006396},
 primaryClass = {astro-ph},
       adsurl = {https://ui.adsabs.harvard.edu/abs/2000AJ....120.1579Y}
}

@ARTICLE{Steinmetz2006,
       author = {{Steinmetz}, M. and {Zwitter}, T. and {Siebert}, A. and {Watson}, F.~G. and {Freeman}, K.~C. and {Munari}, U. and {Campbell}, R. and {Williams}, M. and {Seabroke}, G.~M. and {Wyse}, R.~F.~G. and {Parker}, Q.~A. and {Bienaym{\'e}}, O. and {Roeser}, S. and {Gibson}, B.~K. and {Gilmore}, G. and {Grebel}, E.~K. and {Helmi}, A. and {Navarro}, J.~F. and {Burton}, D. and {Cass}, C.~J.~P. and {Dawe}, J.~A. and {Fiegert}, K. and {Hartley}, M. and {Russell}, K.~S. and {Saunders}, W. and {Enke}, H. and {Bailin}, J. and {Binney}, J. and {Bland-Hawthorn}, J. and {Boeche}, C. and {Dehnen}, W. and {Eisenstein}, D.~J. and {Evans}, N.~W. and {Fiorucci}, M. and {Fulbright}, J.~P. and {Gerhard}, O. and {Jauregi}, U. and {Kelz}, A. and {Mijovi{\'c}}, L. and {Minchev}, I. and {Parmentier}, G. and {Pe{\~n}arrubia}, J. and {Quillen}, A.~C. and {Read}, M.~A. and {Ruchti}, G. and {Scholz}, R.-D. and {Siviero}, A. and {Smith}, M.~C. and {Sordo}, R. and {Veltz}, L. and {Vidrih}, S. and {von Berlepsch}, R. and {Boyle}, B.~J. and {Schilbach}, E.},
        title = "{The Radial Velocity Experiment (RAVE): First Data Release}",
      journal = {\aj},
         year = 2006,
        month = oct,
       volume = {132},
       number = {4},
        pages = {1645-1668},
          doi = {10.1086/506564},
archivePrefix = {arXiv},
       eprint = {astro-ph/0606211},
 primaryClass = {astro-ph},
       adsurl = {https://ui.adsabs.harvard.edu/abs/2006AJ....132.1645S}
}

@ARTICLE{Yanny2009,
       author = {{Yanny}, Brian and {Rockosi}, Constance and {Newberg}, Heidi Jo and {Knapp}, Gillian R. and {Adelman-McCarthy}, Jennifer K. and {Alcorn}, Bonnie and {Allam}, Sahar and {Allende Prieto}, Carlos and {An}, Deokkeun and {Anderson}, Kurt S.~J. and {Anderson}, Scott and {Bailer-Jones}, Coryn A.~L. and {Bastian}, Steve and {Beers}, Timothy C. and {Bell}, Eric and {Belokurov}, Vasily and {Bizyaev}, Dmitry and {Blythe}, Norm and {Bochanski}, John J. and {Boroski}, William N. and {Brinchmann}, Jarle and {Brinkmann}, J. and {Brewington}, Howard and {Carey}, Larry and {Cudworth}, Kyle M. and {Evans}, Michael and {Evans}, N.~W. and {Gates}, Evalyn and {G{\"a}nsicke}, B.~T. and {Gillespie}, Bruce and {Gilmore}, Gerald and {Nebot Gomez-Moran}, Ada and {Grebel}, Eva K. and {Greenwell}, Jim and {Gunn}, James E. and {Jordan}, Cathy and {Jordan}, Wendell and {Harding}, Paul and {Harris}, Hugh and {Hendry}, John S. and {Holder}, Diana and {Ivans}, Inese I. and {Ivezi{\v{c}}}, {\v{Z}}eljko and {Jester}, Sebastian and {Johnson}, Jennifer A. and {Kent}, Stephen M. and {Kleinman}, Scot and {Kniazev}, Alexei and {Krzesinski}, Jurek and {Kron}, Richard and {Kuropatkin}, Nikolay and {Lebedeva}, Svetlana and {Lee}, Young Sun and {French Leger}, R. and {L{\'e}pine}, S{\'e}bastien and {Levine}, Steve and {Lin}, Huan and {Long}, Daniel C. and {Loomis}, Craig and {Lupton}, Robert and {Malanushenko}, Olena and {Malanushenko}, Viktor and {Margon}, Bruce and {Martinez-Delgado}, David and {McGehee}, Peregrine and {Monet}, Dave and {Morrison}, Heather L. and {Munn}, Jeffrey A. and {Neilsen}, Jr., Eric H. and {Nitta}, Atsuko and {Norris}, John E. and {Oravetz}, Dan and {Owen}, Russell and {Padmanabhan}, Nikhil and {Pan}, Kaike and {Peterson}, R.~S. and {Pier}, Jeffrey R. and {Platson}, Jared and {Re Fiorentin}, Paola and {Richards}, Gordon T. and {Rix}, Hans-Walter and {Schlegel}, David J. and {Schneider}, Donald P. and {Schreiber}, Matthias R. and {Schwope}, Axel and {Sibley}, Valena and {Simmons}, Audrey and {Snedden}, Stephanie A. and {Allyn Smith}, J. and {Stark}, Larry and {Stauffer}, Fritz and {Steinmetz}, M. and {Stoughton}, C. and {SubbaRao}, Mark and {Szalay}, Alex and {Szkody}, Paula and {Thakar}, Aniruddha R. and {Sivarani}, Thirupathi and {Tucker}, Douglas and {Uomoto}, Alan and {Vanden Berk}, Dan and {Vidrih}, Simon and {Wadadekar}, Yogesh and {Watters}, Shannon and {Wilhelm}, Ron and {Wyse}, Rosemary F.~G. and {Yarger}, Jean and {Zucker}, Dan},
        title = "{SEGUE: A Spectroscopic Survey of 240,000 Stars with g = 14-20}",
      journal = {\aj},
         year = 2009,
        month = may,
       volume = {137},
       number = {5},
        pages = {4377-4399},
          doi = {10.1088/0004-6256/137/5/4377},
archivePrefix = {arXiv},
       eprint = {0902.1781},
 primaryClass = {astro-ph.GA},
       adsurl = {https://ui.adsabs.harvard.edu/abs/2009AJ....137.4377Y}
}

@PROCEEDINGS{HIPPARCOS,
        title = "{The HIPPARCOS and TYCHO catalogues. Astrometric and photometric star catalogues derived from the ESA HIPPARCOS Space Astrometry Mission}",
    booktitle = {ESA Special Publication},
         year = 1997,
       editor = {{ESA}},
       series = {ESA Special Publication},
       volume = {1200},
        month = jan,
       adsurl = {https://ui.adsabs.harvard.edu/abs/1997ESASP1200.....E}
}

@ARTICLE{Newberg2002,
       author = {{Newberg}, Heidi Jo and {Yanny}, Brian and {Rockosi}, Connie and {Grebel}, Eva K. and {Rix}, Hans-Walter and {Brinkmann}, Jon and {Csabai}, Istvan and {Hennessy}, Greg and {Hindsley}, Robert B. and {Ibata}, Rodrigo and {Ivezi{\'c}}, Zeljko and {Lamb}, Don and {Nash}, E. Thomas and {Odenkirchen}, Michael and {Rave}, Heather A. and {Schneider}, D.~P. and {Smith}, J. Allyn and {Stolte}, Andrea and {York}, Donald G.},
        title = "{The Ghost of Sagittarius and Lumps in the Halo of the Milky Way}",
      journal = {\apj},
         year = 2002,
        month = apr,
       volume = {569},
       number = {1},
        pages = {245-274},
          doi = {10.1086/338983},
archivePrefix = {arXiv},
       eprint = {astro-ph/0111095},
 primaryClass = {astro-ph},
       adsurl = {https://ui.adsabs.harvard.edu/abs/2002ApJ...569..245N}
}

@ARTICLE{Yanny2003,
       author = {{Yanny}, Brian and {Newberg}, Heidi Jo and {Grebel}, Eva K. and {Kent}, Steve and {Odenkirchen}, Michael and {Rockosi}, Connie M. and {Schlegel}, David and {Subbarao}, Mark and {Brinkmann}, Jon and {Fukugita}, Masataka and {Ivezic}, {\v{Z}}eljko and {Lamb}, Don Q. and {Schneider}, Donald P. and {York}, Donald G.},
        title = "{A Low-Latitude Halo Stream around the Milky Way}",
      journal = {\apj},
         year = 2003,
        month = may,
       volume = {588},
       number = {2},
        pages = {824-841},
          doi = {10.1086/374220},
archivePrefix = {arXiv},
       eprint = {astro-ph/0301029},
 primaryClass = {astro-ph},
       adsurl = {https://ui.adsabs.harvard.edu/abs/2003ApJ...588..824Y}
}

@ARTICLE{Majewski2004,
       author = {{Majewski}, Steven R. and {Ostheimer}, James C. and {Rocha-Pinto}, Helio J. and {Patterson}, Richard J. and {Guhathakurta}, Puragra and {Reitzel}, David},
        title = "{Detection of the Main-Sequence Turnoff of a Newly Discovered Milky Way Halo Structure in the Triangulum-Andromeda Region}",
      journal = {\apj},
         year = 2004,
        month = nov,
       volume = {615},
       number = {2},
        pages = {738-743},
          doi = {10.1086/424586},
archivePrefix = {arXiv},
       eprint = {astro-ph/0406221},
 primaryClass = {astro-ph},
       adsurl = {https://ui.adsabs.harvard.edu/abs/2004ApJ...615..738M}
}

@ARTICLE{Martin2007,
       author = {{Martin}, Nicolas F. and {Ibata}, Rodrigo A. and {Irwin}, Mike},
        title = "{Galactic Halo Stellar Structures in the Triangulum-Andromeda Region}",
      journal = {\apjl},
         year = 2007,
        month = oct,
       volume = {668},
       number = {2},
        pages = {L123-L126},
          doi = {10.1086/522791},
archivePrefix = {arXiv},
       eprint = {astro-ph/0703506},
 primaryClass = {astro-ph},
       adsurl = {https://ui.adsabs.harvard.edu/abs/2007ApJ...668L.123M}
}

@ARTICLE{Xu2015,
       author = {{Xu}, Yan and {Newberg}, Heidi Jo and {Carlin}, Jeffrey L. and {Liu}, Chao and {Deng}, Licai and {Li}, Jing and {Sch{\"o}nrich}, Ralph and {Yanny}, Brian},
        title = "{Rings and Radial Waves in the Disk of the Milky Way}",
      journal = {\apj},
         year = 2015,
        month = mar,
       volume = {801},
       number = {2},
          eid = {105},
        pages = {105},
          doi = {10.1088/0004-637X/801/2/105},
archivePrefix = {arXiv},
       eprint = {1503.00257},
 primaryClass = {astro-ph.GA},
       adsurl = {https://ui.adsabs.harvard.edu/abs/2015ApJ...801..105X}
}

@ARTICLE{Schonrich2018,
       author = {{Sch{\"o}nrich}, Ralph and {Dehnen}, Walter},
        title = "{Warp, waves, and wrinkles in the Milky Way}",
      journal = {\mnras},
         year = 2018,
        month = aug,
       volume = {478},
       number = {3},
        pages = {3809-3824},
          doi = {10.1093/mnras/sty1256},
archivePrefix = {arXiv},
       eprint = {1712.06616},
 primaryClass = {astro-ph.GA},
       adsurl = {https://ui.adsabs.harvard.edu/abs/2018MNRAS.478.3809S}
}

@ARTICLE{Kawata2018,
       author = {{Kawata}, Daisuke and {Baba}, Junichi and {Ciuc{\v{a}}}, Ioana and {Cropper}, Mark and {Grand}, Robert J.~J. and {Hunt}, Jason A.~S. and {Seabroke}, George},
        title = "{Radial distribution of stellar motions in Gaia DR2}",
      journal = {\mnras},
         year = 2018,
        month = sep,
       volume = {479},
       number = {1},
        pages = {L108-L112},
          doi = {10.1093/mnrasl/sly107},
archivePrefix = {arXiv},
       eprint = {1804.10175},
 primaryClass = {astro-ph.GA},
       adsurl = {https://ui.adsabs.harvard.edu/abs/2018MNRAS.479L.108K}
}

@ARTICLE{Antoja2022,
       author = {{Antoja}, T. and {Ramos}, P. and {L{\'o}pez-Guitart}, F. and {Anders}, F. and {Bernet}, M. and {Laporte}, C.~F.~P.},
        title = "{Tidally induced spiral arm wraps encoded in phase space}",
      journal = {\aap},
         year = 2022,
        month = dec,
       volume = {668},
          eid = {A61},
        pages = {A61},
          doi = {10.1051/0004-6361/202244064},
archivePrefix = {arXiv},
       eprint = {2206.03495},
 primaryClass = {astro-ph.GA},
       adsurl = {https://ui.adsabs.harvard.edu/abs/2022A&A...668A..61A}
}

@ARTICLE{GaiaCollaboration2021,
       author = {{Gaia Collaboration} and {Antoja}, T. and {McMillan}, P.~J. and {Kordopatis}, G. and {Ramos}, P. and {Helmi}, A. and {Balbinot}, E. and {Cantat-Gaudin}, T. and {Chemin}, L. and {Figueras}, F. and {Jordi}, C. and {Khanna}, S. and {Romero-G{\'o}mez}, M. and {Seabroke}, G.~M. and {Brown}, A.~G.~A. and {Vallenari}, A. and {Prusti}, T. and {de Bruijne}, J.~H.~J. and {Babusiaux}, C. and {Biermann}, M. and {Creevey}, O.~L. and {Evans}, D.~W. and {Eyer}, L. and {Hutton}, A. and {Jansen}, F. and {Klioner}, S.~A. and {Lammers}, U. and {Lindegren}, L. and {Luri}, X. and {Mignard}, F. and {Panem}, C. and {Pourbaix}, D. and {Randich}, S. and {Sartoretti}, P. and {Soubiran}, C. and {Walton}, N.~A. and {Arenou}, F. and {Bailer-Jones}, C.~A.~L. and {Bastian}, U. and {Cropper}, M. and {Drimmel}, R. and {Katz}, D. and {Lattanzi}, M.~G. and {van Leeuwen}, F. and {Bakker}, J. and {Casta{\~n}eda}, J. and {De Angeli}, F. and {Ducourant}, C. and {Fabricius}, C. and {Fouesneau}, M. and {Fr{\'e}mat}, Y. and {Guerra}, R. and {Guerrier}, A. and {Guiraud}, J. and {Jean-Antoine Piccolo}, A. and {Masana}, E. and {Messineo}, R. and {Mowlavi}, N. and {Nicolas}, C. and {Nienartowicz}, K. and {Pailler}, F. and {Panuzzo}, P. and {Riclet}, F. and {Roux}, W. and {Sordo}, R. and {Tanga}, P. and {Th{\'e}venin}, F. and {Gracia-Abril}, G. and {Portell}, J. and {Teyssier}, D. and {Altmann}, M. and {Andrae}, R. and {Bellas-Velidis}, I. and {Benson}, K. and {Berthier}, J. and {Blomme}, R. and {Brugaletta}, E. and {Burgess}, P.~W. and {Busso}, G. and {Carry}, B. and {Cellino}, A. and {Cheek}, N. and {Clementini}, G. and {Damerdji}, Y. and {Davidson}, M. and {Delchambre}, L. and {Dell'Oro}, A. and {Fern{\'a}ndez-Hern{\'a}ndez}, J. and {Galluccio}, L. and {Garc{\'\i}a-Lario}, P. and {Garcia-Reinaldos}, M. and {Gonz{\'a}lez-N{\'u}{\~n}ez}, J. and {Gosset}, E. and {Haigron}, R. and {Halbwachs}, J.-L. and {Hambly}, N.~C. and {Harrison}, D.~L. and {Hatzidimitriou}, D. and {Heiter}, U. and {Hern{\'a}ndez}, J. and {Hestroffer}, D. and {Hodgkin}, S.~T. and {Holl}, B. and {Jan{\ss}en}, K. and {Jevardat de Fombelle}, G. and {Jordan}, S. and {Krone-Martins}, A. and {Lanzafame}, A.~C. and {L{\"o}ffler}, W. and {Lorca}, A. and {Manteiga}, M. and {Marchal}, O. and {Marrese}, P.~M. and {Moitinho}, A. and {Mora}, A. and {Muinonen}, K. and {Osborne}, P. and {Pancino}, E. and {Pauwels}, T. and {Recio-Blanco}, A. and {Richards}, P.~J. and {Riello}, M. and {Rimoldini}, L. and {Robin}, A.~C. and {Roegiers}, T. and {Rybizki}, J. and {Sarro}, L.~M. and {Siopis}, C. and {Smith}, M. and {Sozzetti}, A. and {Ulla}, A. and {Utrilla}, E. and {van Leeuwen}, M. and {van Reeven}, W. and {Abbas}, U. and {Abreu Aramburu}, A. and {Accart}, S. and {Aerts}, C. and {Aguado}, J.~J. and {Ajaj}, M. and {Altavilla}, G. and {{\'A}lvarez}, M.~A. and {{\'A}lvarez Cid-Fuentes}, J. and {Alves}, J. and {Anderson}, R.~I. and {Varela}, E. Anglada and {Audard}, M. and {Baines}, D. and {Baker}, S.~G. and {Balaguer-N{\'u}{\~n}ez}, L. and {Balog}, Z. and {Barache}, C. and {Barbato}, D. and {Barros}, M. and {Barstow}, M.~A. and {Bartolom{\'e}}, S. and {Bassilana}, J.-L. and {Bauchet}, N. and {Baudesson-Stella}, A. and {Becciani}, U. and {Bellazzini}, M. and {Bernet}, M. and {Bertone}, S. and {Bianchi}, L. and {Blanco-Cuaresma}, S. and {Boch}, T. and {Bombrun}, A. and {Bossini}, D. and {Bouquillon}, S. and {Bragaglia}, A. and {Bramante}, L. and {Breedt}, E. and {Bressan}, A. and {Brouillet}, N. and {Bucciarelli}, B. and {Burlacu}, A. and {Busonero}, D. and {Butkevich}, A.~G. and {Buzzi}, R. and {Caffau}, E. and {Cancelliere}, R. and {C{\'a}novas}, H. and {Carballo}, R. and {Carlucci}, T. and {Carnerero}, M.~I. and {Carrasco}, J.~M. and {Casamiquela}, L. and {Castellani}, M. and {Castro-Ginard}, A. and {Castro Sampol}, P. and {Chaoul}, L. and {Charlot}, P. and {Chiavassa}, A. and {Cioni}, M.-R.~L. and {Comoretto}, G. and {Cooper}, W.~J. and {Cornez}, T. and {Cowell}, S. and {Crifo}, F. and {Crosta}, M.},
        title = "{Gaia Early Data Release 3. The Galactic anticentre}",
      journal = {\aap},
         year = 2021,
        month = may,
       volume = {649},
          eid = {A8},
        pages = {A8},
          doi = {10.1051/0004-6361/202039714},
archivePrefix = {arXiv},
       eprint = {2101.05811},
 primaryClass = {astro-ph.GA},
       adsurl = {https://ui.adsabs.harvard.edu/abs/2021A&A...649A...8G}
}

@ARTICLE{Friske2019,
       author = {{Friske}, Jennifer K.~S. and {Sch{\"o}nrich}, Ralph},
        title = "{More than just a wrinkle: a wave-like pattern in U$_{g}$ versus L$_{z}$ from Gaia data}",
      journal = {\mnras},
         year = 2019,
        month = dec,
       volume = {490},
       number = {4},
        pages = {5414-5423},
          doi = {10.1093/mnras/stz2951},
archivePrefix = {arXiv},
       eprint = {1902.09569},
 primaryClass = {astro-ph.GA},
       adsurl = {https://ui.adsabs.harvard.edu/abs/2019MNRAS.490.5414F}
}

\begin{appendix}
\section{Spatial distribution of the sample}

\begin{figure*}[t]
\sidecaption
\includegraphics[width=12cm]{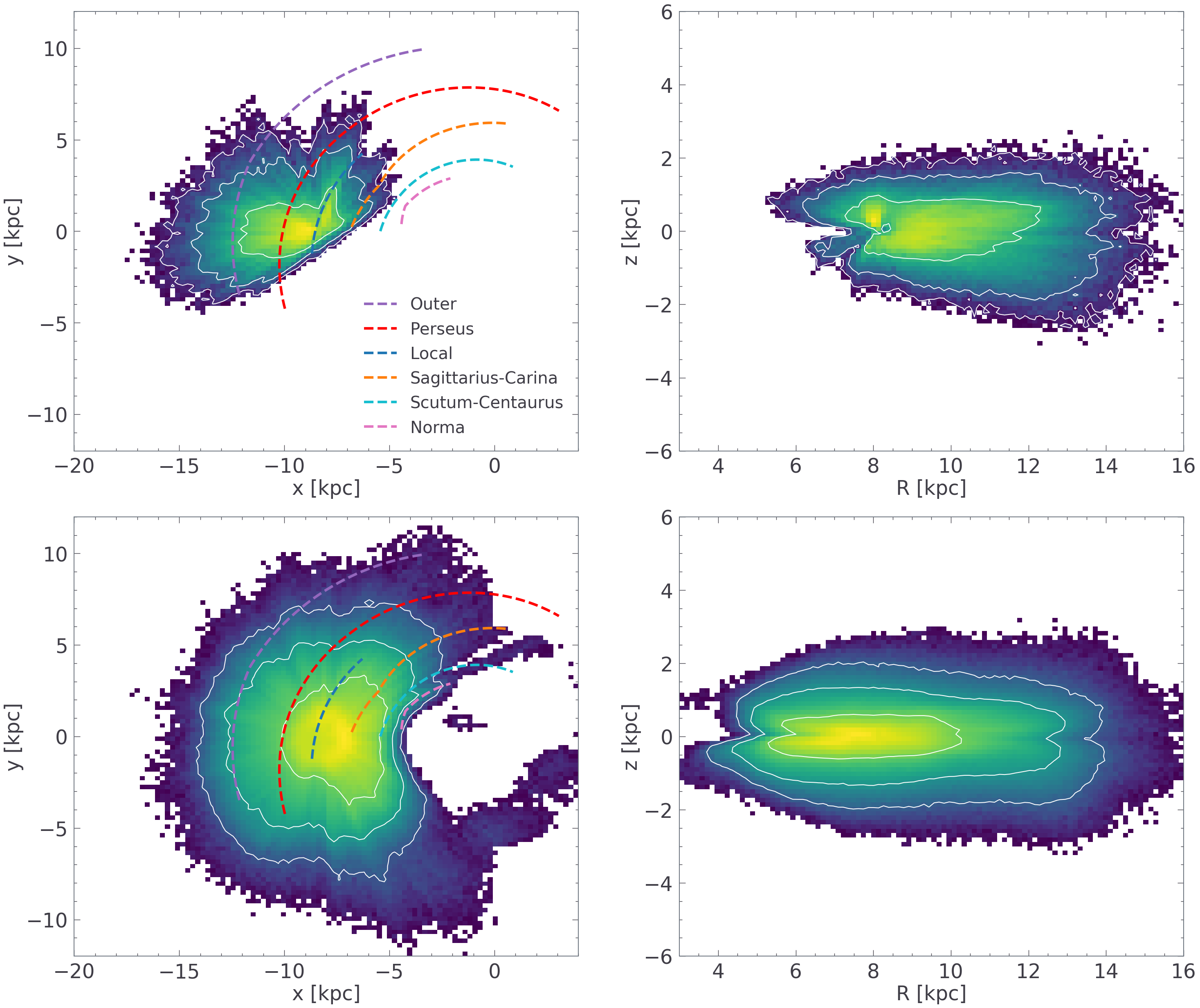}
\caption{Spatial distribution of the disk sample from LAMOST (first row) and Gaia (second row). Number-density maps are shown in the $x$--$y$ and $R$--$z$ planes. Bins with fewer than five stars are omitted. The white contours indicate the regions enclosing 68\%, 95\%, and 99\% of the stars. Spiral arm models from \citet{reid2019trigonometric} are overplotted in all panels.}
\label{fig:spatial_spiral_arms}
\end{figure*}

As shown in Fig.~\ref{fig:spatial_spiral_arms}, our sample spatially overlaps with six spiral arms. The spiral arm loci are adopted from \citet{reid2019trigonometric}, which are based on trigonometric parallax measurements of molecular masers. 

\section{N-body settings and methods}\label{sec:Nbody}

The particles used in the N-body simulations are generated by a more realistic MW model constructed by DFs via AGAMA package \citep{Vasiliev2019}. In equilibrium, the DF can be taken as a function of the three action variables $(J_R,J_z,J_\phi)$, which are isolating integrals of the motion \citep{Binney2008}, where $J_R$ quantifies the amplitude of radial excursions, $J_z$ oscillations away from the Galactic plane, and $J_\phi$ is the component of angular momentum parallel to the Galaxy’s assumed symmetry axis. Following the modelling work in \cite{Binney2023}, an equilibrium MW model can be defined by a superposition of similar DFs. The DF of each Galactic component is a specified function $f(\vJ)$ of the action integrals. 

\begin{figure*}
 \includegraphics[width=\textwidth]{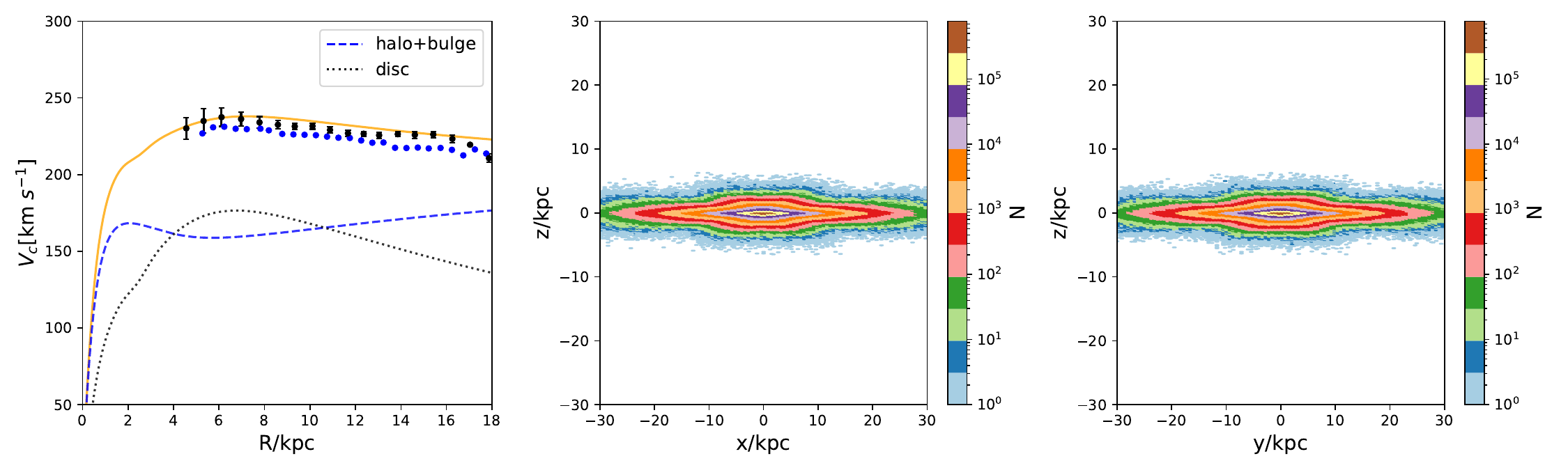}
 \caption{Circular velocity and distributions in $x-z$, and $y-z$ plans of the initial disk particles. In the left panel, the orange curve shows the total circular velocity curve of the Galaxy derived by the DF model, the dashed blue and dotted curves denote the spheroidal and disk components respectively. The black and blue data points are the observational results based on Ceipheids \citep{Ablimit2020} and red giants \citep{Eilers2019} respectively. The middle and right panels show the spatial distribution of the initial disk particles, in $x-z$ and $y-z$ plane respectively. }
 \label{fig:circular_velocity}
\end{figure*}

In this model, the disk components have exponential DFs, which is physically reasonable throughout action space compared to quasi-isothermal DFs. A double power law DF is used to model the bulge, stellar and dark halo. The total gravitational potential of the MW is then derived from the DFs. The details of the descriptions of this MW axisymmetric model can be found in \cite[]{Binney2023,Binney2024}. The circular velocity generated from this axisymmetric potential is shown in the upper panel of Fig.~\ref{fig:circular_velocity}.

The spheroidal components in the MW are constructed using modified double power-law DFs exactly the same as those in \cite{Binney2023}, including the bulge and the dark and stellar halos.
\begin{equation}
    f(\vJ)=\frac{M}{(2\pi J_{0})^{3}}\frac{(1+[J_{0}/h_{J}]^{\gamma})^{\alpha_{\rm{in}}/\gamma}}
    {(1+[g_{J}/J_{0}]^{\gamma})^{\alpha_{\rm{out}}/\gamma}}e^{-(g_{J}/J_{cut})^{\delta}},
    \label{eq:db}
\end{equation}

The modifications of this model compared with \cite{Li2022a} are the replacement of the linear combinations of $g(\vJ)$ and $h(\vJ)$, which endeavours to solve the unphysical behaviour velocity distributions at small $\lvert \rm{v}_\phi \rvert$. 
In this model, the $g(\vJ)$ and $h(\vJ)$ are:
\begin{equation}
    g(\vJ) = h(\vJ) = J_R \rm e^{-\beta \sin{(c\pi/2)}}
    + \frac{1}{2}(1+c\xi) \rm e^{\beta \sin{(c\pi/2)}} \mathcal{L},
    \label{eq:gh}
\end{equation}
where 

\begin{equation}
 c \equiv \frac{L}{L+J_R},
 \label{eq:c}
\end{equation}
is the measurement of the orbital circularity and $L=\lvert J_\phi \rvert +J_z$ is the angular momentum. 
The realization of the total angular momentum reads,
\begin{equation}
 \mathcal{L}(J_z,J_\phi) \equiv \rm a J_z + \rm b
 \frac{\lvert J_\phi \rvert J_z}{L} + \lvert J_\phi \rvert, 
 \label{eq:Lvar}
\end{equation}
where the constants $\rm a$ and $\rm b$ are chosen to ensure that the value of $\frac{\partial{\mathcal{L}}}{\partial{v_\phi}}$ vanishes at $v_\phi\rightarrow0$, which deals with the unphysical behaviour at small $\lvert \rm{v}_\phi \rvert$. The constants $\rm a$ and $\rm b$ are determined by:
\begin{equation}
 a = \frac{1}{2}(k+1)\quad \rm and \quad b = \frac{1}{2}(k-1), 
 \label{eq:ab}
\end{equation}
where the value of $k$ is determined by 
\begin{equation}
 k(\xi) = (1-\xi) \rm F_{in} + \xi \rm F_{out} 
 \label{eq:k}
\end{equation}
$\rm F_{in}$ and $\rm F_{out}$ are two constant numbers that listed in Table~\ref{tab:df_sph} for the halo and bulge. The parameter $\xi$ is another representative of the energy which is defined as 
\begin{equation}
 \xi \equiv \frac{j_t^\alpha}{j_t^\alpha + 1},
 \label{eq:xi}
\end{equation}
where $\alpha=0.6$ in our work and 
\begin{equation}
 j_t \equiv \frac{1.5J_R+L}{L_0},
 \label{eq:jt}
\end{equation}
where $L_0$ is a scale action of $L_0=6J_0$.
The last term in Equation~\ref{eq:db} measures the cutoff of the bulge and halo components, in which $J_{cut}$ is the cutoff action and $\delta$ is the cutoff strength.

This disk model follows \cite{Binney2024}, which adopted truncated exponential DF models for the disks as:
\begin{equation}
    \label{eq:disc}
f(\vJ)=f_\phi(J_\phi)f_r(J_R,J_\phi)f_z(J_z,J_\phi)f_{\rm{ext}}(J_\phi)f_{\rm{int}}(J_\phi).
\end{equation}

In this model, the function $f_r$ controls the velocity dispersions $\sigma_R$ and $\sigma_\phi$ near the Galactic plane. The function $f_z$ controls both the thickness of the disk and the velocity dispersion $\sigma_z$. The factor $f_\phi$ generates a roughly exponentially declining surface density $\Sigma(R)\simeq\exp(-R/R_\d)$. The factor $f_{\rm ext}$ and $f_{\rm int}$ truncates the disk at some outer and inner radius respectively. 
The young thin disk behaves taper exponential profile which owns a truncated radius in the inner disk following the work from \cite{Li2022b}. The high-$\alpha$ disk favours a sharpness density decline in large radii at about $J_\phi \approx 2000 \kms \kpc$. 
The factors $f_{\rm ext}$ and $f_{\rm int}$ are:

\begin{equation}\label{intext}
\begin{split}
    &f_{\rm{int}}(J_\phi)= \frac{e^{2(J_\phi-J_{\rm int})/D_{\rm int}}}{e^{2(J_\phi-J_{\rm int})/D_{\rm int}}+1},\\
    &f_{\rm{ext}}(J_\phi)=\frac{e^{-2(J_\phi-J_{\rm ext})/D_{\rm ext}}}{e^{-2(J_\phi-J_{\rm ext})/D_{\rm ext}}+1},
\end{split}
\end{equation}
where the action $J_{\rm int}$ and $J_{\rm ext}$ determines the characteristic radius of the central
depression or outer cutoff. The parameters $D_{\rm int}$ and $D_{\rm ext}$ controls the sharpness of the transitions. The function for $f_R(\vJ)$ and $f_z(\vJ)$ are
\begin{equation}\label{frz}
\begin{split}
    &f_{\rm{R}}(\vJ)= \frac{1}{J_{\rm R0}}  \left( \frac{J_v}{J_{\phi0}} \right)^{p_R} \exp{\left[-\frac{1}{J_{\rm R0}}  \left( \frac{J_v}{J_{\phi0}} \right)^{p_R} J_R \right]},\\
    &f_{\rm{z}}(\vJ)= \frac{1}{J_{\rm z0}}  \left( \frac{J_v}{J_{\phi0}} \right)^{p_z} \exp{\left[-\frac{1}{J_{\rm z0}}  \left( \frac{J_v}{J_{\phi0}} \right)^{p_z} J_z \right]},
\end{split}
\end{equation}
where $J_{\phi0}$ is the action that correlates with the disk’s scale length and the action $J_{R0}$ and $J_{z0}$ determine the velocity dispersions $\sigma_R$ and $\sigma_z$. $p_R$ and $p_z$ control the velocity dispersion profile along the Galactic radial direction.
$J_{v} \equiv J_{R} + J_z + J_\phi + J_{v0}$, where $J_{v0}$ is a scale action that controls the velocity dispersion at small radii. The function form of $J_\phi$ is  
\begin{equation}\label{eq:jphi}
f_\phi(\vJ)=\frac{\rm M}{(2\pi)^3} \frac{J_d}{J^2_{\phi0}} e^{-J_d/J_{\phi0}},
\end{equation}
where $J_{d} \equiv J_{R} + J_z + J_\phi + J_{d0}$, and $J_{d0}$ is a scale action that controls the surface density at small radii. The detailed description and the predictions of this model can be found in \cite{Binney2024}. The DF parameters for the disk and spheroidal components are listed in  Table~\ref{tab:df} and Table~\ref{tab:df_sph} respectively. The orange curve in the left panel in Fig.~\ref{fig:circular_velocity} shows the total circular velocity curve of the Galaxy derived by the DF model, the dashed blue and dotted curves denote the spheroidal and disk components respectively. The black and blue data points are the observational results based on Ceipheids \citep{Ablimit2020} and red giants \citep{Eilers2019} respectively. The middle and right panels of  Fig.~\ref{fig:circular_velocity} shows the spatial distribution of the initial disk particles, in $x-z$ and $y-z$ plane respectively. There are neither warp nor flare signatures in the initial disk.

The other preparation is the computing of the initial position and velocity for the Sgr. The initial position of the Sgr 2 Gyr ago is estimated by integrating backwards the current position and velocity of Sgr, given by $\mathbf{x}_{\rm Sgr,0} = \{17.9, 2.6, -6.6\}\ \rm{kpc}$ and $\mathbf{v}_{\rm Sgr,0} = \{239.5, -29.6, 213.5\}\ \rm{km\ s^{-1}}$ \citep{Vasiliev21}, under the axisymmetric MW model, which is computed by the DF models through a self-consistently iterating procedure \citep{Binney2014}. At 2 Gyr ago, the Sgr is modelled as a Plummer profile with a mass of $2 \times 10^{9}\ M_{\odot}$ and a scale radius $r_{\rm s}$ of 1.5 kpc. The dynamical friction experienced by the Sgr during the backward orbital integration is approximated using Chandrasekhar’s standard formula \citep{Chandrasekhar1943}:

\begin{align}
    \mathbf{a}_{\rm DF} &= -4\pi G^2 \rho_{\rm MW}(\mathbf{x}_{\rm Sgr}) M_{\rm Sgr} \ln\Lambda \nonumber \\
    &\quad \times \frac{\mathbf{v}_{\rm Sgr}}{|\mathbf{v}_{\rm Sgr}|^3} \left[ \operatorname{erf}(X) - \frac{2X}{\sqrt{\pi}} e^{-X^2} \right],
\end{align}

where \( \rho_{\rm MW}(\mathbf{x}_{\rm Sgr}) \) denotes the density of the MW halo at the position of Sgr, and the Coulomb logarithm is fixed at \( \ln\Lambda = 3 \) following \citet{Vasiliev21}. The parameter \( X \) is defined as \( X = \frac{|\mathbf{v}_{\rm Sgr}|}{\sqrt{2} \sigma_{\rm MW}} \), where \( \sigma_{\rm MW} \) is the local velocity dispersion of the MW halo \citep{Vasiliev21}.

A constant mass-loss rate of 0.2\% per 0.01~Gyr is adopted for Sgr, resulting in a cumulative mass loss of approximately 70\% over 2~Gyr. The integrated initial position of Sgr 2~Gyr ago is listed in Table~\ref{tab:sgr_ini}. 

The $N$-body simulations of TDH and Sgr were carried out using \textsc{Gadget-4} \citep{Springel2021}. The initial conditions of particles for the MW components and Sgr were generated from their DF models using \textsc{Agama}.

The particle mass of each MW component and Sgr is determined by dividing the total mass of the component by the number of particles, with the resulting value rounded to the nearest integer. The particle numbers and masses for each component and Sgr are listed in Table~\ref{tab:nbody}.

We adopt an empirical formula for the gravitational softening length, which depends solely on the particle mass \citep{Moster14}:
\begin{equation}\label{eq:epsilon} 
\epsilon = 32 \sqrt{\frac{M_p}{10^{10} M_\odot}},
\end{equation}
where $M_p$ is the mass of an individual particle in solar masses.

\begin{table*}
\footnotesize
    \begin{center}
    \caption{Parameters used for the DF model of the disk components.}
    \label{tab:df}
    \begin{tabular*}{\textwidth}{@{\extracolsep{\fill}}lcccccccccccc}
        \hline
        $\rm Name$ &$M$ &$J_{\phi0}$ &$J_{R0}$ &$J_{z0}$ &$J_{\rm ext}$ &$D_{\rm ext}$ &$J_{\rm int}$ &$D_{\rm int}$ &$p_r$ &$p_z$ &$J_{\rm v0}$ &$J_{\rm d0}$ \\ 
        \hline 
        $\rm Young$  &$0.45$ &$977.9$ &$2.806$ &$1.296$ &$--$ &$--$ &$186.9$ &$278$  &$-0.76$ &$-0.23$ &$152.1$ &$102.1$ \\
        $\rm Mid$  &$1.2$ &$1030$ &$22.82$ &$3.24$ &$--$ &$--$ &$--$ &$--$ &$-0.23$ &$-0.7$ &$146.4$ &$731.9$ \\
        $\rm Old$  &$1.47$ &$508$ &$47.14$ &$10.04$ &$--$ &$--$ &$--$ &$--$ &$0.034$ &$-0.043$ &$132.7$ &$150$ \\
        $\rm Thick$  &$1.3$ &$399$ &$116.4$ &$54.6$ &$2212$ &$207.8$ &$--$ &$--$ &$0.1$ &$0.17$ &$150$ &$10$ \\
        \hline
    \end{tabular*}
    \end{center}
\tablefoot{
The units of mass and actions are $10^{10}\,M_\odot$ and km\,s$^{-1}$\,kpc, respectively.
$J_{\phi0}$ determines the radial profile of the disk components.
$J_{R0}$ and $J_{z0}$ determine the velocity dispersion profiles in the $R$ and $z$ directions.
$p_r$ and $p_z$ control the sharpness of the velocity dispersion profiles.
$D_{\rm ext}$, $J_{\rm ext}$ and $D_{\rm int}$, $J_{\rm int}$ determine the outer and inner truncations, respectively.
The parameter values are based on the fit to APOGEE stars by \citet{Binney2024}.
}
\end{table*}

\begin{table*}
\begin{center}
    \caption{Parameters used for the spheroidal DF model of the bulge and halo.}
    \label{tab:df_sph}
        \begin{tabular*}{\textwidth}{@{\extracolsep{\fill}}lcccccccccc}
            \hline
            $\rm Name$ & $M$ & $\alpha_{\rm in}$ & $\alpha_{\rm out}$ & $J_{\rm core}$ & $J_{\rm cut}$ & $J_{0}$ & $F_{\rm in}$ & $F_{\rm out}$ & $\beta$ & $\delta$\\ 
            \hline 
            $\rm Bulge$ & $0.1$ & $0.5$ & $1.8$	& $5$ & $200$ & $19.5$ & $3$ & $2$ & $0.6$ & $2$\\
            $\rm Stellar\,Halo$ & $0.04$ & $1.6$ & $4.2$ & $6.148$ & $4e+04$	& $583.2$ & $1.8$ & $1.2$ & $0.5$ & $2$\\
            $\rm Dark\,Halo$ & $65$ & $1.6$ & $2.7$ & $100$ & $2e+04$	& $10000$ & $1.4$ & $1.2$ & $0$ & $2$\\
            \hline
        \end{tabular*}
\end{center}
\tablefoot{The units of mass and actions are $10^{10}\msun$ and $\kms\kpc$ respectively. $\alpha_{\rm in}$ and $\alpha_{\rm out}$ represent the inner and outer slopes in the double power law model. $J_{\rm cut}$ denotes the outer cut of the halo and bulge and $\delta$ defines the sharpness of the downturn. $J_0$ is the break action which determines the break radius of the inner and outer part. $F_{\rm in}$, $F_{\rm out}$, and $\beta$ determine the shape of the components and the velocity anisotropy.}
\end{table*}

\begin{table*}
\begin{center}
    \caption{Initial phase space coordinates of the Sgr dwarf galaxy in the N-body simulation.}
    \label{tab:sgr_ini}
        \begin{tabular*}{0.6\linewidth}{cccccc}
            \hline
             $x$ & $y$ & $z$ & $v_x$ & $v_y$ & $v_z$\\ 
             $(\rm kpc)$ & $(\rm kpc)$ & $(\rm kpc)$ & $(\rm km\ s^{-1})$ & $(\rm km\ s^{-1})$ & $(\rm km\ s^{-1})$\\ 
            \hline 
            $-6.97$ & $3.53$ & $-15.00$	& $277.08$ & $36.47$ & $-210.19$\\
            \hline
        \end{tabular*}
\end{center}
\end{table*}

\begin{table*}
\centering
\small
\caption{Particle numbers and particle masses for each component used in the N-body simulations.}
\label{tab:nbody}
\begin{tabularx}{\textwidth}{@{\extracolsep{\fill}}lXXXXXXXX}
\hline
Component & Bulge & Young Disk & Middle Disk & Old Disk & Thick Disk & Stellar Halo & Dark Halo & Sgr \\ 
\hline 
Particle Number ($10^6$) & $5$ & $6$ & $13$ & $13$ & $13$ & $0.025$ & $75$ & $0.5$ \\
Particle Mass $(\rm M_{\odot})$ & $1526$ & $710$ & $765$ & $1054$ & $888$ & $12188$ & $14635$ & $4000$ \\
\hline
\end{tabularx}
\end{table*}

\end{appendix}

\end{CJK*}
\end{document}